\documentclass[fleqn,usenatbib]{mnras}

\usepackage{newtxtext,newtxmath}
\usepackage[T1]{fontenc}

\DeclareRobustCommand{\VAN}[3]{#2}
\let\VANthebibliography\thebibliography
\def\thebibliography{\DeclareRobustCommand{\VAN}[3]{##3}\VANthebibliography}

\usepackage{graphicx}	% Including figure files
\usepackage{amsmath}	% Advanced maths commands

\usepackage{amssymb}	% Extra maths symbols
\usepackage{xspace}
\usepackage[table]{xcolor}
\usepackage[english]{babel} % To obtain English text with the blindtext package
\usepackage{blindtext}
\usepackage[export]{adjustbox}

\graphicspath{{plot/}}
\newcommand{\thethreehundred}{{\sc The Three Hundred}}
\newcommand{\GIZ}{\textsc{Gizmo-SIMBA}\xspace}
\newcommand{\GX}{\textsc{Gadget-X}\xspace}
\newcommand{\ahf}{\textsc{AHF}\xspace}
\newcommand{\Newton}{\emph{XMM-Newton}\xspace}
\newcommand{\Chandra}{\emph{Chandra}\xspace}
\newcommand{\eROSITA}{\emph{eROSITA}\xspace}
\newcommand{\Planck}{\emph{Planck}\xspace}

\usepackage[normalem]{ulem}
\title[Baryon profiles]{\thethreehundred\ Project: the evolution of physical baryon profiles}

\author[Q. Li et al.]{\parbox{\textwidth}{
Qingyang Li,$^{1,2,5}$\thanks{E-mail: qingyli@sjtu.edu.cn}
Weiguang Cui,$^{3,4,5}$\thanks{E-mail: weiguang.cui@roe.ac.uk; Talento-CM fellow}
Xiaohu Yang,$^{1,2,6}$ Romeel Dav$\mathrm{\acute{e}}$,$^5$ 
Elena Rasia,$^{7,8}$ Stefano Borgani,$^{7,8,9,10}$ Meneghetti Massimo,$^{11}$ Alexander Knebe,$^{3,4,12}$
Klaus Dolag$^{13,14}$ and Jack Sayers$^{15}$
}
\vspace{0.4cm}
\\
\parbox{\textwidth}{
$^1$Department of Astronomy, School of
  Physics and Astronomy, Shanghai Jiao Tong University, Shanghai 200240, China\\
$^2$Shanghai Key Laboratory for Particle Physics and Cosmology, Shanghai 200240, China\\
$^{3}$Departamento de Física Teórica, M-8, Universidad Autónoma de Madrid, Cantoblanco 28049, Madrid, Spain\\
$^{4}$Centro de Investigación Avanzada en Física Fundamental (CIAFF), Universidad Aut\'{o}noma de Madrid, Cantoblanco, 28049 Madrid, Spain\\
$^5$Institute for Astronomy, University of Edinburgh, Royal Observatory, Edinburgh EH9 3HJ, United Kingdom\\
$^6$Tsung-Dao Lee Institute, Shanghai Jiao Tong University, Shanghai 200240, China\\
$^{7}$IFPU, Institute for Fundamental Physics of the Universe, Via Beirut 2, 34014 Trieste, Italy\\
$^{8}$INAF -- Osservatorio Astronomico di Trieste, via Tiepolo 11, I-34131, Trieste, Italy \\
$^{9}$Dipartimento di Fisica dell'Universit\`a di Trieste, Sez. di Astronomia, via Tiepolo 11, I-34131 Trieste, Italy \\
$^{10}$INFN, Instituto Nazionale di Fisica Nucleare, Via Valerio 2, I-34127, Trieste, Italy\\
$^{11}$INAF—OAS, Osservatorio di Astrofisica e Scienza dello Spazio di Bologna, via Gobetti 93/3, I-40129 Bologna, Italy\\
$^{12}$International Centre for Radio Astronomy Research, University of Western Australia, 35 Stirling Highway, Crawley, Western Australia 6009, Australia\\
$^{13}$Universitäts-Sternwarte, Fakult$\ddot{a}$t f$\ddot{u}$r Physik, Ludwig-Maximilians-Universit$\ddot{a}$t M$\ddot{u}$nchen, Scheinerstr. 1, 81679 M$\ddot{u}$nchen, Germany\\
$^{14}$Max-Planck-Institut f$\ddot{u}$r Astrophysik, Karl-Schwarzschild-Stra$\mathcal{\beta}$e 1, D-85741 Garching, Germany\\
$^{15}$California Institute of Technology, 1200 East California Boulevard, Pasadena, California 91125, USA
}}
\date{Accepted XXX. Received YYY; in original form ZZZ}

\pubyear{2023}

\begin{document}
\label{firstpage}
\pagerange{\pageref{firstpage}--\pageref{lastpage}}
\maketitle

% Abstract of the paper
\begin{abstract}
The distribution of baryons provides a significant way to understand the formation of galaxy clusters by revealing the details of its internal structure and changes over time. In this paper, we present theoretical studies on the scaled profiles of physical properties associated with the baryonic components, including gas density, temperature, metallicity, pressure and entropy as well as stellar mass, metallicity and satellite galaxy number density in galaxy clusters from $z=4$ to $z=0$ by tracking their progenitors. These mass-complete simulated galaxy clusters are coming from \thethreehundred\ with two runs: \GIZ\ and \GX. Through comparisons between the two simulations, and with observed profiles which are generally available at low redshift, we find that (1) the agreements between the two runs and observations are mostly at outer radii $r \gtrsim 0.3r_{500}$, in line with the self-similarity assumption. While \GX\ shows better agreements with the observed gas profiles in the central regions compared to \GIZ; (2) the evolution trends are generally consistent between the two simulations with slightly better consistency at outer radii. 
% Gas density, metallicity and pressure are increasing with redshift moving to 0, while temperature and entropy decrease with redshift. 
In detail, the gas density profile shows less discrepancy than the temperature and entropy profiles at high redshift. The differences in the cluster centre and gas properties imply different behaviours of the AGN models between \GX\ and \GIZ, with the latter, maybe too strong for this cluster simulation. The high-redshift difference may be caused by the star formation and feedback models or hydrodynamics treatment, which requires observation constraints and understanding.
% (3) The satellite number density profiles in both simulations reasonably match well with low-redshift observations at the outskirts.
\end{abstract}

% Select between one and six entries from the list of approved keywords.
% Don't make up new ones.

%%%%%%%%%%%%%%%%%%%%%%%%%%%%%%%%%%%%%%%%%%%%%%%%%%

%%%%%%%%%%%%%%%%% BODY OF PAPER %%%%%%%%%%%%%%%%%%

\section{Introduction}
As the largest gravitationally bound structures in the Universe, galaxy clusters provide the unique environment to study galaxy formation and cosmology \citep[e.g.,][]{Yang2007,Allen2011,KS2012,Rykoff2014,Yang2021,Li2022b}. 
Although the baryonic components including intracluster medium (ICM) and galaxies occupy a small fraction of cluster mass, they are the significant ways to investigate the properties and history of clusters in observation \citep[e.g.,][]{Tozzi2001,Cavagnolo2008}, and to deduce the amount of total matter by scaling relation \citep[e.g.,][]{Bialek2001,Giodini2013}. Besides, baryon distribution in galaxy clusters also provides the constraints on the physical models describing chemical enrichment \citep[e.g.,][]{Scannapieco2005} and the astrophysical processes like active galactic nuclei (AGN) feedback \citep[][]{Teyssier2011,Martizzi2012,Sembolini2016} and gas cooling \citep[e.g.,][]{Eckert2012,Li2020}.

The ICM in galaxy clusters is mostly contributed by hot ionized plasma which has been successfully detected by X-ray telescopes (e.g., \Newton, \Chandra and \eROSITA). In addition, the distribution of ICM can also be traced by the Sunyaev-Zel'dovich (SZ) effect \citep{SZ1972},
% {Planck2013,Romero2017,Ruppin2018,Romero2020}, 
which is produced by the inverse Compton scattering of the Cosmic Microwave Background (CMB) photons with hot electrons in the ICM. Large galaxy cluster samples have been determined through SZ sky surveys such as the Atacama Cosmology Telescope \citep[ACT;][]{ACT2021}, the South Pole Telescope \citep[SPT;][]{SPT2015} and \Planck \citep{Planck2016b}. Based on these abundant observational data, the distribution of ICM has been fully investigated with different physical quantities mostly at low redshifts, including mass density \citep[e.g.,][]{Vikhlinin2006,Ettori2013}, temperature \citep[e.g.,][]{Vikhlinin2005,Pratt2007}, pressure \citep[e.g.,][]{Finoguenov2007,Arnaud2010}, entropy \citep[e.g.,][]{Pratt2010,Mantz2016} and metallicity \citep[e.g.,][]{Baldi2007,Leccardi2008b,Lovisari2019}. At high redshifts, limited by telescope resolution and detection sensitivity, the radial profiles of only a few far-away individual clusters can be investigated with joint observational analysis. For example, \citet{Ruppin2020b} presented a multi-wavelength analysis of the massive cluster MOOJ1142+1527 at $z = 1.2$ using high angular resolution \Chandra X-ray and NIKA2 SZ data \citep[see also other works, such as ][]{Hilton2010,Tozzi2015,Brodwin2016,Keruzore2020,Sayers2023}. In particular, the profiles of these physical properties in galaxy clusters generally present a self-similarity at large radii after deducting evolution effects \citep[e.g.,][]{Leccardi2008,Ghirardini2019}.

On the other hand, archival cluster samples at different redshifts allow us to compare cluster property changes, more interestingly with ICM physical profiles, using X-ray observations \citep[e.g.,][]{Baldi2012, Ghirardini2017,Bulbul2019} or SZ surveys \citep[e.g.,][]{McDonald2013,McDonald2016,Ruppin2020}. Many statistical studies showed that the ICM profiles evolved with redshifts still hold the self-similarity \citep[e.g.,][]{Baldi2012b,McDonald2017,Bartalucci2017}, especially in the outskirts, but a substantial amount of scatters in the core due to the inclusion of a fraction of clusters with a cool and dense core.
\citet{McDonald2014} studied the evolution of mean temperature, pressure and entropy profiles with 80 galaxy clusters selected from $z=0$ to $z=1.2$ in the SPT survey and \Chandra X-ray observation. Their pressure profiles matched with previous low-redshift measurements, showing little redshift evolution in the outer region. \citet{Ettori2015} presented the analysis of spatially-resolved metal abundance with a sample of 83 galaxy clusters in the redshift range 0.09-1.39 using \Newton data. They found no evidence of redshift evolution of the metallicity at $r > 0.4r_{500}$\footnote{The radius, $r_{500}$, is defined as the location in clusters where the average matter density within this radius is 500 times the critical density of the universe.} but a negative evolution in the inner region. \citet{Sanders2018} reported a self-similar evolution of the thermodynamic properties at all radii for the \Chandra observed sample but using a centre of X-ray peak and a slightly different analysis scheme out to $r_{500}$. Furthermore, the statistical analysis of the evolution of hot plasma properties has extended out to $z \sim 1.9$ lately. For example, \citet{McDonald2017} presented the self-similar ICM density profiles in the eight massive galaxy clusters selected from the SPT catalogue at $1.2<z<1.9$. \citet{Ghirardini2021} studied the thermodynamic properties including density, temperature, pressure and entropy with seven massive and distant clusters ($1.2<z<1.8$) selected from ACT observations and compared with lower-redshift samples ($z<0.1$). They found that both high and low redshift clusters show similar profiles in the outskirts following a self-similar model \citep{Kaiser1986}. 

However, note that these property changes among clusters at different redshifts are not necessarily their evolution, i.e., the observed clusters at high redshift are not necessarily the progenitors of the low redshift clusters \citep[see][for example]{Cui2020}. The redshift evolution (or no evolution) could be due to the sample selection, for instance, a redshift invariant halo mass cut. The advantage of hydrodynamic simulation is that the true progenitors of clusters are known, which allows us to investigate the profile changes along the cluster formation and evolution.
We focus on a census of the evolution of physical ICM profiles from $z=4$ to $z=0$ and explore how the self-similarity performs and holds even at the early period of cluster forming and out to $r_{500}$. Furthermore, we compare two versions of hydrodynamic simulations (\GIZ and \GX, which have different baryon models and are tuned to represent different observation properties), which allow us to understand how the different baryon models, especially the non-gravitational processes, such as AGN feedback in the centre of clusters, affect the distribution of ICM. 

The distribution of ICM in the simulated galaxy clusters (see these projects: MUSIC \citep{Sembolini2013}, Rhapsody-G \citep{Hahn2017}, MACSIS \citep{Barnes2017a}, Hydrangea \citep{Bahe2017}, FABLE \citep{Henden2018} and ROMULUSC \citep{Tremmel2019}, for example) is usually studied by an overall investigation on one specific property \citep[e.g.,][]{Vogelsberger2018, Biffi2018} or general presentation of several properties at $z = 0$ \citep[e.g.,][]{Barnes2017b, Henden2018, Tremmel2019, Pakmor2022}. \cite{Pearce2021} focused on the evolution of  metallicity abundance out to $z = 2$ using C-EAGLE clusters. However, the inconsistency with observations for some physical properties still exists in simulations. \citet{Oppenheimer2021} discussed the commons and differences of baryon distribution in different simulations. Using the EAGLE-like simulations, \citet{Altamura2022} found that the high entropy in the cores of clusters can be eliminated without a run of artificial conduction, metal cooling, or AGN feedback but still fails to match with observation.

Besides, we are interested in theoretical investigations on the evolution of stellar and satellite galaxies profiles, i.e. the final results of star formation, which allows us to probe the effects of the implemented baryon models. In observations, stellar profiles are mainly observed at optical bands \citep[e.g.,][]{van2015,Cariddi2018}. \citet{Annunziatella2014} determined the stellar mass density profile of a cluster of galaxies -- MACS J1206.2-0847 -- at $z = 0.44$ selected from the CLASH-VLT survey \citep{Postman2012}. Their result showed a decreasing radial trend of the average stellar mass. Meanwhile, the investigation of the profiles of stellar components becomes more accurate, as the ongoing and upcoming optical surveys, such as Hyper Suprime-Cam Subaru Strategic Program \citep[HSC-SSP,][]{HSCPDR3} and Legacy Survey of Space and Time \citep[LSST,][]{LSST}, and missions carried by \emph{Euclid} space telescope \citep{Euclid}, will dramatically increase the image quality and resolution of cluster samples.

The radial distribution of cluster galaxies has been widely investigated in observations \citep[e.g.,][]{Carlberg1997,Guo2012,Presotto2012,Hartley2015,Shin2021}, especially for the number density which is found to be well described by a simple power-law \citep[e.g.,][]{Sales2005,Chen2006,Lares2011} or the Navarro-Frenk-White~\citep[NFW;][]{NFW} model \citep[e.g.,][]{Muzzin2007}. \citet{Hartley2015} found the slope of fitted power-law in a range $-1.1$ to $-1.4$ for the stellar mass selected satellites, and $-1.3$ to $-1.6$ for passive satellites using a deep near-infrared dataset. \citet{Budzynski2012} found that the shape of radial galaxy number density presents no strong variation with redshift and exhibits a high degree of self-similarity after applying a projected NFW model to SDSS7 data. Most of the observations focus on low or intermediate redshifts. However, theoretically understanding the satellite number density profile at high redshifts ($z > 2$) helps us to interpret the formation of clusters at early times -- the protocluster regions.

In Section~\ref{sec:data}, we introduce the simulation data with the two baryon models. The evolution of ICM properties profiles including mass density, electron number density, temperature, entropy, pressure and metallicity is presented in Section~\ref{sec:gas}. We discuss the distribution of stellar and satellite galaxies in Section~\ref{sec:stellar}. The conclusions are summarized in Section~\ref{sec:cons}. Throughout the paper, we adopt the cosmology parameters from \emph{Planck} 2016 results \citep{Planck2016}: $\Omega_\mathrm{m} = 0.307$, $\Omega_{\Lambda} = 0.693$ and $h = H_0/(100\ \rm km\ s^{-1}\ Mpc^{-1)}=0.678$.

%------------------------------------------------------------------------

\section{Data} \label{sec:data}

\subsection{Hydrodynamical simulations}

The two hydrodynamic simulations named \GIZ \citep{Cui2022} and \GX \citep{Rasia2015} from \thethreehundred\ Project\footnote{\url{https://the300-project.org}}~\citep[hereafter the300,][]{Cui2018} produced 324 galaxy clusters starting from the same initial conditions. These clusters are selected from the $N$-body simulation of MultiDark Planck 2~\citep[MDPL2,][]{Klypin2016}\footnote{\url{https://www.cosmosim.org/cms/simulations/mdpl2}}. Then, using the re-generated initial conditions, \GIZ and \GX are run with the same \Planck\ 2016 cosmology parameters as the parent simulation~\citep{Planck2016}. The high-resolution cluster zoom regions, centred on the 324 most massive clusters at $z=0$ in MDPL2 which are identified by the \textsc{rockstar}~\citep{Behroozi2012} halo finder\footnote{\url{https://bitbucket.org/gfcstanford/rockstar}}, have a radius of 15 $h^{-1}\mathrm{Mpc}$. The dark matter (DM) and gas-particle masses in the high-resolution region are $m_{\rm DM}\simeq 12.7 \times 10^8\ h^{-1}\mathrm{M_{\odot}}$ and $m_{\rm gas} \simeq 2.36\ \times 10^8\ h^{-1}\mathrm{M_{\odot}}$. \GIZ\ and \GX\ set fixed softening lengths to values of $5\ h^{-1}\rm kpc$ and $6.5\ h^{-1}\rm kpc$, respectively. We briefly illustrate the details of baryon models for \GX\ and \GIZ\ in the following subsection. More details can be found in \citet{Dave2016,Dave2019,Cui2022} (for \GIZ) and \citet{Cui2018} (for \GX).
Benefiting from the unique setups, these simulated clusters from the300 project have been widely used for different studies, for example, environment effect \citep{WangYang2018}, cluster profiles \citep{Mostoghiu2019, Li2020, Baxter2021}, splash-back galaxies \citep{Arthur2019, Haggar2020, Knebe2020}, cluster dynamical state \citep{DeLuca2021, Capalbo2021, Zhang2021, Li2022}, filament structures \citep{Kuchner2020, Rost2021, Kuchner2021}, lensing studies \citep{Vega-Ferrero2021, Herbonnet2022, Giocoli2023}, cluster mass \citep{Li2021, Gianfagna2023} and machine learning studies \citep{deAndres2022, deAndres2023, Ferragamo2023}.

\subsection{Baryon models}

%%%GIZMO
\GIZ\ runs by the \textsc{Gizmo} code \citep{Hopkins2015} with a Meshless Finite Mass (MFM) to solve hydrodynamic components. The MFM solver improves the description of shocks and flows with high Mach numbers and the handling of contact discontinuities relative to smoothed particle hydrodynamics (SPH). The state-of-the-art galaxy formation baryon models are similar to that in the recent \textsc{Simba} simulation \citep{Dave2019}. The implementation of gas radiative cooling and photon-heating/ionization process is based on the \texttt{Grackle-3.1} library \citep{Smith2017}. An $\rm H_2$-based star formation model is adopted from the predecessor simulation \textsc{MUFASA} \citep{Dave2016}. The stellar feedback models \citep{Muratov2015,Alcazar2017} adopt two-phase winds with mass loading factor scaling with the galaxy stellar mass identified by an on-fly friends-of-friends (FoF) finder.
The chemical enrichment model tracks eleven elements from SNIa, SNII and AGB stars. Besides, the metal of wind particles is subtracted from nearby gas in a kernel-weighted manner. \textsc{Simba} \citep{Dave2019} includes two models of black hole (BH) growth prescriptions: the torque-limited accretion model from cold gas \citep{Alcazar2015,Alcazar2017b}, and the Bondi accretion from hot gas. AGN feedback is modelled by kinetic bipolar outflows, the strength of which depends on the black-hole accretion rate, separated into three modes: a "radiative mode" at high Eddington ratio to drive multi-phase winds at velocities of $\sim 10^3\ \rm s^{-1}$km, a "jet mode" at low Eddington ratios ($f_{\rm Edd} < 0.2$), at which AGNs drive hot gas in collimated jets at velocities of $\sim10^4\ \rm s^{-1}$km, and X-ray heating from black holes, which aims to represent the momentum input from hard photons radiated off from the accretion disk \citep{Choi2012}, and which is only initiated after jet-mode feedback is turned on. We note that the parameters of the \GIZ\ model are re-calibrated compared with the original \GIZ\ simulation \citep{Dave2019} in order to suit the lower mass resolution of the300 clusters and match observations well. The recalibration is based on observed stellar properties: total stellar mass fraction within $r_{500}$, BCG stellar mass - halo mass relation, and the satellite galaxy stellar mass function in galaxy clusters. We give a brief introduction about this recalibration in the following \citep[more details in ][]{Cui2022}. The calibration is done with a single cluster region at $z = 0$ where the largest object has $M_{500} \approx 5 \times 10^{14}\ \rm M_{\odot}$, selected from the300 sample. Using a single sample avoids spending an infeasible amount of computation time. In detail, we strengthened the AGN jet feedback by increasing the maximum jet speed which overcomes the insufficient star formation in satellite galaxies and higher stellar mass fraction within $r_{500}$ caused by low mass resolution. We used a fixed comoving softening length to be consistent with other the300 runs. We also increased the BH accretion kernel maximum radius and the BH seeding stellar mass. Once the recalibration based on this region was done to match observations sufficiently, all parameters were frozen and run for all the remaining galaxy clusters.

%%%Gadget-X
\GX\ is based on the gravity solver of the {\sc GADGET3} Tree-PM code (an updated version of the {\sc GADGET2} code; \citealt{Springel2005}). The SPH scheme is improved with artificial thermal diffusion, time-dependent artificial viscosity, a high-order Wendland C4 interpolating kernel and a weak-up scheme \citep{Beck2016}. Stellar evolution and metal enrichment \citep[see][for the original formulation]{Tornatore2007} consider mass-dependent lifetimes of stars \citep{Padovani1993}, the production and evolution of 15 different elements coming from SNIa, SNII and AGB stars with metallicity-dependent radiative cooling \citep{Wiersma2009}. The stellar feedback model is adopted from \citet{Springel2003} with a wind velocity of 350 $\rm s^{-1} km$. The \GX\ models also include a BH growth and implementation of AGN feedback \citep{Steinborn2015}. The BH accretion is based on the Eddington-limited Bondi accretion \citep{Bondi1952} with individual accretion rates for hot (boost factor $\alpha = 10$) and cold ($\alpha = 100$) gas respectively, while the AGN feedback, modelling both the mechanical and radiative modes, is implemented as thermal feedback.

While the gravity and hydro solvers in simulations accurately generate underlying gravitational framework \citep{Sembolini2016}, the main differences in baryon distribution between \GX\ and \GIZ\ are caused by the adopted baryon models. As \GX\ is successful in reproducing the observational gas distribution \citep{Li2020} and \GIZ\ is primarily tuned to reproduce galaxy stellar properties, thus the emphasis on the adjustment for different baryon distribution in two hydros will help us to understand better how to model galaxy formation in galaxy clusters.

\subsection{Halo sample and profile calculation}

The haloes in these re-simulation regions are identified by the Amiga Halo Finder~\citep[\ahf,][]{Knollmann2009} with an overdensity of 200 $\times\ \rm \rho_{crit}$, where $\rm \rho_{crit}$ is the critical density of the universe at the corresponding redshift.
% We first select the mass-complete the300 samples at $z$ = 0, i.e., central clusters.
The main progenitors of the central clusters are identified by tracing the merger tree, which is determined by the {\sc MERGERTREE} package integrated into the AHF program. We define the main progenitor as the most massive halo among all progenitors in the previous snapshot. More detailed information on these clusters and their progenitors can be found in Table~\ref{tab:1}.

\begin{table*}
    \centering
    \caption{Halo mass ranges of progenitors at different redshifts. Row (a) lists the median of halo mass, $\log M_{500}$, in a unit of $\rm M_{\odot}$, while row (b) lists the 16th and 84th percentiles of $\log M_{500}$. Row (c) indicates the number of progenitors at each redshift. Note that the two numbers at $z=4$ indicate the progenitors after excluding these with few gas and stellar particles, respectively.} 
    \label{tab:1}
    \begin{tabular}
    {c c c c c c c}
    \hline \hline
        & & $z = 0$ & $z = 1$ & $z = 2$ & $z = 3$ & $z = 4$\\
    %   & & ($10^{14}\ h^{-1}\rm M_{\odot}$) & ($10^{14}\ h^{-1}\rm M_{\odot}$) & ($10^{13}\ h^{-1}\rm M_{\odot}$) & ($10^{13}\ h^{-1}\rm M_{\odot}$)\\
     \hline
      \GIZ\ & (a) & 14.91 & 14.17 & 13.47 & 12.91 & 12.39\\
      & (b) & [14.78, 15.04] & [13.87, 14.48] & [13.11, 13.84] & [12.51, 13.32] & [11.93, 12.83] \\
      & (c) & 324 & 324 & 323 & 323 & 317, 309
    \\
    \\
      \GX\ & (a) & 14.93 & 14.19 & 13.49 & 12.92 & 12.40 \\
      & (b) & [14.81, 15.06] & [13.86, 14.50] & [13.14, 13.86] & [12.52, 13.33] & [11.98, 12.82] \\
      & (c) & 324 & 324 & 324 & 323 & 302, 309\\
    \hline
    \end{tabular}
\end{table*}

For the profiles in this paper, we focus on the differential profiles, i.e. the corresponding values at each radius. In detail, we only use hot gas for the gas profiles: temperature $T > 10^6\ \rm K$, delay time $t_{\rm d} \le 0$ \footnote{This only applies to \GIZ\ which is to exclude wind particles. The delay time marks the time after which the wind particles can be decoupled and treated as normal particles.} and gas density $\rho < 2.88 \times 10^6\ \rm kpc^{-3}M_{\odot}$ in order to compare our results with X-ray observations. Note that the criterion of delay time does not apply to \GX. Both gas and stellar profiles are calculated extending to $1.5r_{200}$ in logarithmic space. At $z = 4$, radial bins are merged with the nearby bins to avoid noise due to fewer particles number in high-redshift progenitors, mostly haloes with low mass. If a halo at the corresponding redshift has a total gas/stellar particle number (after applying the previous hot gas selection criteria) less than 50 within $1.5r_{200}$, we simply remove it from the total sample. Throughout this paper, we calculate the gas and stellar profiles from simulations in three dimensions. However, the number density of satellite galaxies is determined in a projected plane for comparisons with observations.

\section{Gas profiles}  \label{sec:gas}

\subsection{Global heating history for gas in haloes} \label{sec:gas_history}

\begin{figure}
    \centering
    \includegraphics[width=0.48\textwidth]{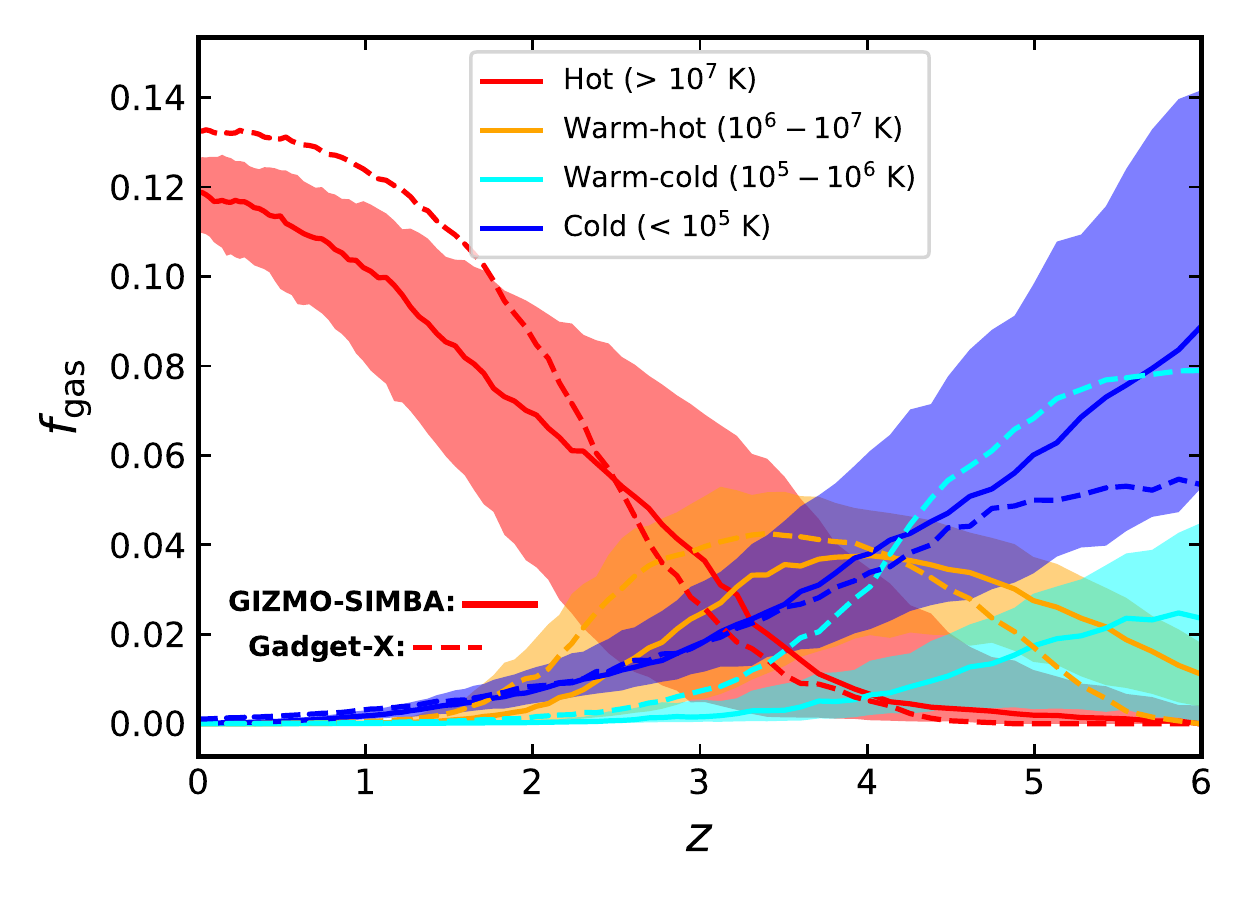}
    \caption{Evolution of gas mass fraction at different temperature phases. The hot ($>10^7$ K) and cold ($<10^5$ K) phases are respectively shown with red and blue colours, while warm-hot and warm-cold phases with temperature range $10^6 - 10^7$ K and $10^5$ -- $10^6$ K are shown with orange and cyan colours, respectively. The solid and dashed lines indicate the median results from \GIZ\ and \GX, while the shadow regions show the 16th and 84th percentiles of all clusters in \GIZ.}
    \label{fig:fgas_evo}
\end{figure}

\begin{figure*}
    \centering
    \includegraphics[width=\textwidth]{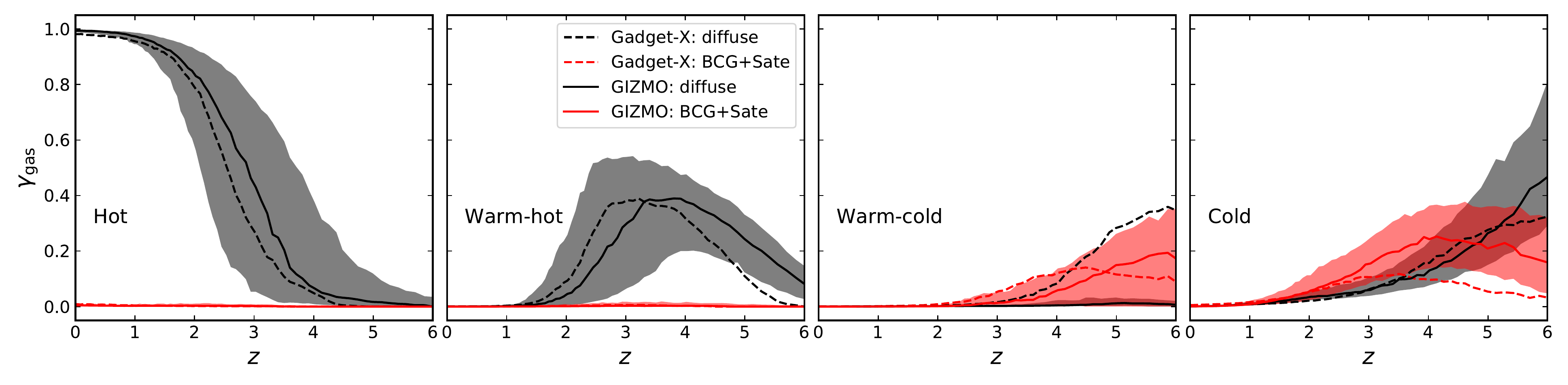}
    \caption{The gas abundances in the bound and diffuse conditions as a function of redshift. From the left to the right panels, these abundances are shown at hot, warm-hot, warm-cold and cold temperature phases, respectively. The solid and dashed lines indicate the evolution in \GIZ\ and \GX, while the black and red lines show gas particles in the bound (BCG and satellite galaxies) and diffuse conditions, respectively. The shaded regions are the 16th and 84th percentiles of all clusters in \GIZ.}
    \label{fig:fsubgas_evo}
\end{figure*} 

Though the gas fraction evolution has been shown in \citet{Cui2022} (see their Fig. 3 for details), it is worth checking how the gas is distributed in different temperature ranges. This gives us a full picture of the gas component within the galaxy clusters at different redshifts and allows us to understand the thermalization history of galaxy clusters \citep{Sereno2021}. In Fig.~\ref{fig:fgas_evo}, we show the evolution of gas mass fraction within the progenitor haloes\footnote{Note here that all the progenitor haloes up to $z=6$ are included in this analysis as we are focusing on the global quantities and an even higher redshift presents a more complete picture of gas thermalization history.} at different temperature phases (hot, warm-hot, warm-cold and cold) as a function of redshift, which comes from 89 snapshots ranging from $z=6$ to $z=0$. The gas mass fraction, $f_{\rm gas}$, is calculated by the ratio between the corresponding gas mass and the total mass within $r_{500}$. The hot, warm and cold phases are defined as the temperature ranges at $> 10^7$ K, $10^5 - 10^7$ K and $< 10^5$ K respectively \citep[e.g.,][for a similar definition]{Cui2019, Li2020}. 
% While the temperature of our gas particle selection is large than $10^6$ K, 
We specifically split the warm phases into two phases, named warm-hot and warm-cold with a temperature range of $10^6 - 10^7$ K and $10^5 - 10^6$ K respectively, in order to make a clear understanding in the following discussion.   

At high redshift, the gas mass fraction is dominated by cold (or cold and warm-cold for \GX) gas in both simulations, which is not surprising. This difference -- more warm-cold gas in \GX\ than in \GIZ\ at high redshifts -- is possibly caused by different cooling models, UV/X-ray background radiation models and SN feedback models \citep[see][for more discussions]{Cui2022}. Besides that difference, both cold and warm-cold gas fractions decrease along redshift, and the fraction differences become no noticeable below $z \sim 2 - 3$. The warm-hot gas fraction increases and crosses the cold gas fraction at $z\sim 4$. After its peak at $z \sim 3-4$, the warm-hot gas fraction decreases and becomes almost negligible after $z \sim 2$. The source of heating for the warm gas probably comes from both the SN feedback, possibly dominating the warm-cold gas phase for there is a clear difference between \GIZ\ and \GX, and the gravitational shock which could be the major contribution to the warm-hot gas because there is less difference between the two simulations and their halo virial temperatures fall into this range. The hot gas fraction only starts to increase after $z \sim 4$, and eventually overtakes all the other fractions at $z \sim 3$, slightly earlier in \GIZ\ than \GX. At this redshift, the median halo mass is around $10^{13}\ \rm M_{\odot}$ -- a critical halo mass for quenching \citep[see e.g.][for the debated discussions]{Zu2016, Cui2021}. Note here that our profile studies stop at $z=4$ because there is less hot gas after $z=4$. The heating of gas could be due to both the AGN feedback and accretion shock. After $z\sim2$, the halo circumgalactic medium (CGM) is dominated ($\gtrsim 90\%$) by the high-temperature gas. Later accreted gas in these galaxy clusters, especially after $z\sim1$, must be either already in the hot phase or heated up very quickly (shock heating). Although a general picture of the gas evolution in the formation of galaxy clusters is very similar between the two simulations, there are subtle fraction differences. Further, note that the differences in the total gas fractions \citep[see][for detailed gas fraction evolution]{Cui2022} between the two simulations may also have some influences on these evolution differences.

It is also interesting to understand where the gas is located in the galaxy clusters. We explore that by further separating the gas fractions into bound and diffuse conditions: (1) gas particles in the brightest central galaxy (BCG) and satellite galaxies/subhaloes; (2) gas particles in the diffuse environment, respectively. The BCG is defined as the spherical region with a radius of $0.05r_{500}$ at the centre of clusters. This radius roughly corresponds to 72 kpc and 67 kpc by averaging all the clusters in \GIZ\ and \GX\ at $z =0$, respectively. Only the gas particles within that radius are taken as bound gas to the BCG. The gas particles which are bound to subhaloes within $r_{200}$ are identified by \ahf.
% we simply use that information from the \ahf catalogue. 
On the other side, the gas particles lying outside BCGs and satellite galaxies/subhaloes are treated as "diffuse" components. We note that the definition of BCG is only a rough estimation, but this should be enough for us to roughly understand how baryons are distributed inside galaxy clusters. Careful studies on this require a proper definition of BCG \citep[see][for how to disentangle the BCGs from the intra cluster light in hydrosimulation and their strong method dependencies]{Cui2014} and the comparisons to observations should be done with proper mock images, which are beyond the scope of this paper. In Fig.~\ref{fig:fsubgas_evo}, we present the bound and diffuse separations with $\gamma_{\rm gas}$, which is defined as:
\begin{equation}
	\gamma_{\rm gas} = \frac{M_X}{M_{\rm gas}}, \label{eq:gamma}
\end{equation}  
where $M_X$ is the total gas mass of either bound or diffuse particles and $M_{\rm gas}$ is the total gas mass within $r_{500}$. We show the evolution of $\gamma_{\rm gas}$ in the hot, warm-hot, warm-cold and cold temperature phases following Fig.~\ref{fig:fsubgas_evo}. It is clear that the hot and warm-hot gas is hardly ever bounded to galaxies at all redshifts. At the warm-cold phase, a fraction of gas is bounded in both \GIZ and \GX at high redshift $z \gtrsim 3$, while a significant fraction of gas in \GX is reserved in the diffused environment with almost zero in \GIZ. At high redshift, cold gas can be in both bounded and diffuse environments which can be caused by our definition of BCG. There is less bounded cold gas in \GX than that in \GIZ at $z \gtrsim 2$ which explains the high star formation rate in \GIZ seen in \cite{Cui2022} and in Section \ref{sec:stellar}.
      
In the following subsections, we quantify the gas distributions through the radial profiles of different properties, including mass density, electron number density, temperature, entropy, pressure and metallicity. As mentioned before, gas particles are selected with a temperature larger than $10^6$ K, we equivalently call these gas components ICM, which is composed of superheated plasma. This is because we would like to compare the gas profiles with these observational results mostly from X-ray telescopes. We note that the cold gas only contributes a very small faction up to $z\sim 2$ as shown in Fig.~\ref{fig:fgas_evo} \citep[see also][for $z$ = 0 result]{Li2020} and begins to dominate the gas component since $z=4$, hence the overall distribution of gas components can be reflected by ICM in galaxy clusters through our profile shows.

\subsection{Mass density} \label{sec:density}
\begin{figure}
    \centering
    \includegraphics[width=0.48\textwidth]{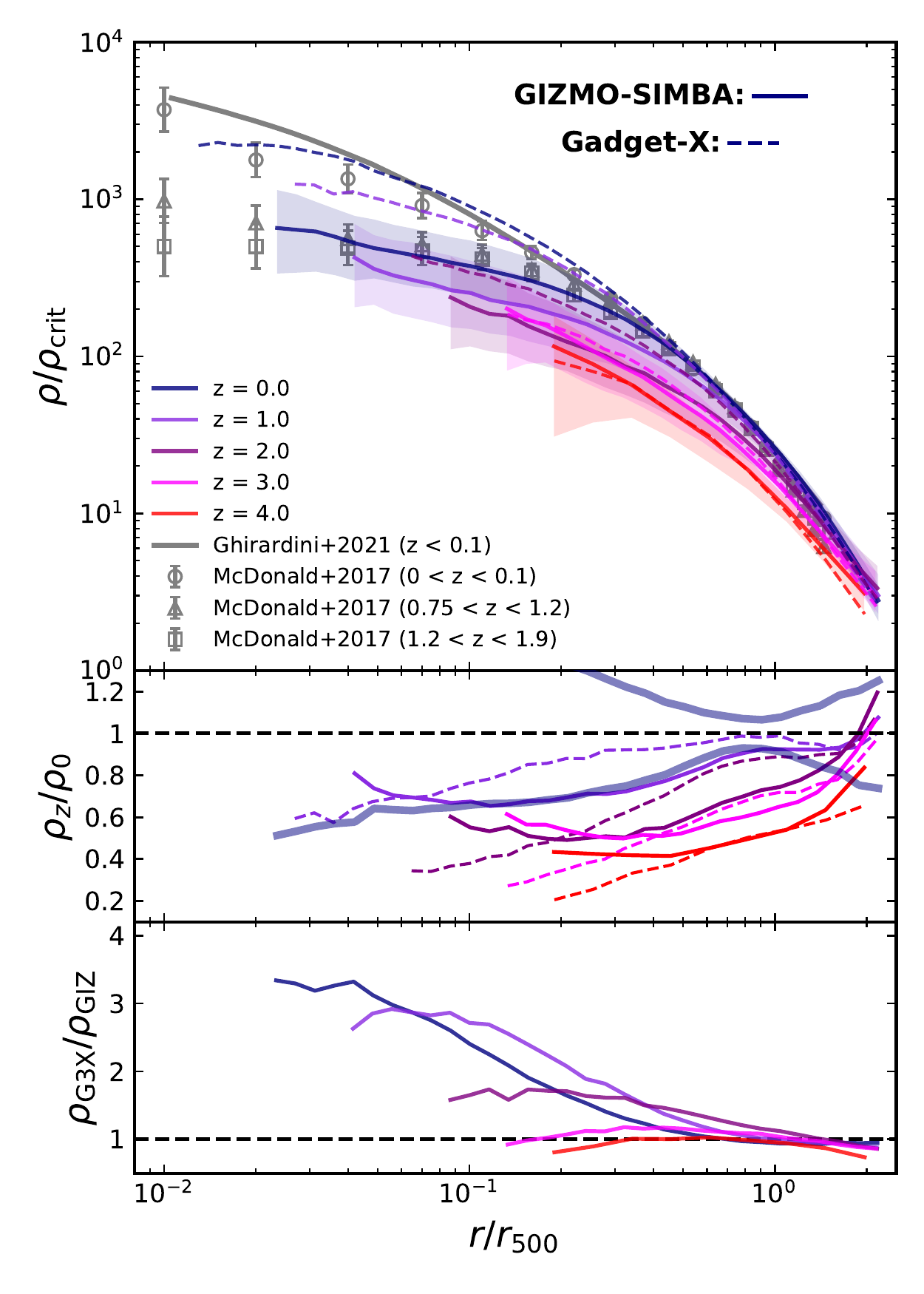}
    \caption{The evolution of gas mass density profiles normalized by critical density of the universe from $z=4$ to $z=0$. The solid and dashed lines in the top panel show the median profiles for \GIZ and \GX with a gradually shallow colour from low redshifts to high redshifts. The shaded regions represent the $16$th and $84$th percentile of all cluster profiles in \GIZ. The hollow grey circles, triangles and squares respectively show the observations in \citet{McDonald2017} at three redshift bins: $0 < z < 0.1$, $0.75 < z < 1.2$ and $1.2 < z < 1.9$. The grey line represents the observations in \citet{Ghirardini2021} at $z < 0.1$. The middle panel shows the ratio between the normalized density profiles at $z \geq 1$ and that at $z = 0$. The bold navy blue line marks the scatter range of the density at $z = 0$. The bottom panel shows the ratio between the density in \GX\ and \GIZ\ at the same redshift.}
    \label{fig:Gdens}
\end{figure}

The previous section only exhibits the overall gas abundance and evolution, while the density profile can further show how gas is distributed within galaxy clusters and it has been used to predict the cluster mass with the hydrostatic equilibrium assumption \citep[see][and references therein]{Li2021,Gianfagna2022}. Though high-order statistics, such as quadrupole \citep{Lokken2022}, may provide even detailed information for physical properties and the model behind \citep[for example,][found the CGM anisotropy is correlated with the collimated AGN feedback]{Yang2023}, we restrict our analysis only to this 1D profile as it is not only a direct and fundamental reflection of gas amount and distribution but also highly correlated with other gas properties. Therefore, we start with the gas density profile.

In the top panel of Fig.~\ref{fig:Gdens}, we present the gas mass density profiles from $z = 4$ to $z = 0$ in \GIZ (solid lines) and \GX (dashed lines). The mass density is calculated along the radial direction, following the formula:
\begin{equation}
    \rho = \frac{1}{V} \sum_i m_{\mathrm{gas}, i},
\end{equation}
where $V$ is the volume of a shell at corresponding bins.
The density is rescaled to the critical density of the universe ($\rho_{\rm crit} \equiv 3H^2(z)/8\pi\mathrm{G}$) at corresponding redshifts, while the radius is normalized to $r_{500}$. We also include observational results for comparisons in this plot. The grey symbols show the observational profiles at $0<z<0.1$, $0.75<z<1.2$ and $1.2<z<1.9$ taken from \citet{McDonald2017}, which constrained the evolution of ICM using selected galaxy clusters from X-ray \citep{Vikhlinin2009} and SZ surveys \citep{Bleem2015}. The range of cluster mass at the three redshift bins is $4\times 10^{14}\ \mathrm{M_{\odot}} < M_{500} < 1.2 \times 10^{15}\ \mathrm{M_{\odot}}$, $M_{500} \sim 4.2\pm 0.2 \times 10^{14}\ \rm M_{\odot}$, and $2\times 10^{14}\ \mathrm{M_{\odot}} < M_{500} < 4 \times 10^{14}\ \mathrm{M_{\odot}}$, respectively. Another observational profile from \citet{Ghirardini2021} is based on a sample of 12 clusters \citep{Ghirardini2019} with $3.48 \times 10^{14}\ \mathrm{M}_{\odot} \leq M_{500} \leq 8.95 \times 10^{14}\ \mathrm{M}_{\odot}$. These clusters were originally selected from the \emph{Planck} SZ catalogue \citep{Planck2014} with an SZ signal limitation and low redshift (0.04 < $z$ < 0.1). Note that our sample covers the same cluster mass range at $z \sim 0$, of which progenitors have a slightly smaller mass than observations.

In the core region, $r \lesssim 0.1r_{500}$, \GX\ presents a higher profile than that in \GIZ at low redshifts. This could be due to the decoupled kinetic AGN feedback model adopted in \GIZ, which ejects too much gas out and results in a small gas fraction within $r_{500}$ as investigated in \citet{Cui2022}. While \GIZ\ applies a pure kinetic scheme to model AGN feedback, and \GX\ uses thermal schemes of its AGN feedback model, we speculate that this kinetic scheme is more efficient to blow away the central gas. Actually, as shown in \cite{Yang2023}, the gas can be ejected out to $\gtrsim 2 \times r_{200}$. However, compared to observational results at $z < 0.1$, \GX\ shows a quite consistent distribution but with a more "cored" centre, which is in line with the results from \citet{Rasia2015}, who found that some of the simulated clusters based on the same \GX\ code, present a cool-core feature similar to observational data. This performance is competitive to the "cuspy" profile shown in observations, which though becomes less peaky at $z \gtrsim 1.2$ \citep{McDonald2013,Ghirardini2021}. In addition, the larger scatters in the inner region are generally thought of as due to the dichotomy between cool-core and non-cool-core states \citep[e.g.,][]{Li2020}. The inner cut of the gas density profile at high-redshift is due to the resolution limitation, i.e., less number of gas particles involved in this radius.  

At $r \gtrsim 0.3r_{500}$, the differences between \GX\ and \GIZ\ vanish and both are consistent with observations. Meanwhile, the density profiles become steeper with a decreasing scatter until $\sim r_{500}$. According to \citet{Kaiser1986}, the evolution of properties of ICM components follows self-similar scaling results. Here, we investigate the self-similarity of ICM density by checking the normalized density ratio between the profiles at $z \geq 1$ and at $z = 0$, while the result is shown in the middle panel of Fig.~\ref{fig:Gdens}. We consider the ratio between the percentile errors and median value of density at $z = 0$ as the criterion for judging self-similarity, which is presented with bold navy blue lines in the middle panel of Fig.~\ref{fig:Gdens} for highlighting the redshift evolution. The density ratio outside this bold line range is regarded as deviating from self-similarity. Although this criterion is not straightforward as that of fitting the radial profiles with an index-free self-similar formula \citep[for example,][]{McDonald2017,Ghirardini2021}, it is a convenient and direct reference. We emphasize that our limitation is rather stricter than the fitting method as the density error is much small, especially in the outskirt. 
Overall, the density profiles perform a self-similar profile below $z \sim 1$ in both hydro-simulations, consistent with observations \citep[e.g.,][]{Vikhlinin2006,Croston2008,Mantz2016}. Furthermore, a strong deviation at the largest radius is shown at $z = 1$. This could be caused by accretion shock bringing in more gas. At $z = 2$, the profile in \GX\ is roughly located at the edge of our self-similar criterion, while the profile in \GIZ\ has begun to show a deviation from self-similarity within $r_{500}$. Besides, the clusters at $z\sim 1.8$ from observational sample are still holding a self-similarity at $r \gtrsim 0.2r_{500}$ \citep{McDonald2017,Ghirardini2021}. This could be caused by the difference in mass range. Then, the deviation becomes large with increasing redshifts. However, we note that the high-redshift objects have a lower mass on average than their descendants at lower redshifts, thus the deviation from self-similarity at high redshifts may be caused by the high sensitivity of low-mass objects to non-gravitational physics, rather than due to changes with redshift. We examined this idea by fixing the cluster mass at $10^{14}\ \mathrm{M_{\odot}} < M_{500} < 10^{14.3}\ \rm M_{\odot}$ as shown in Fig.~\ref{fig:dens_appd} in Appendix~\ref{sec:fixedM}. There are two interesting effects: at inner radii, $r\lesssim 0.4 \times r_{500}$, \GX\ shows almost no redshift evolution, which indicates the density evolution in Fig.~\ref{fig:Gdens} is purely caused by mass evolution in \GX\ up to $z=2$. While \GIZ\ presents clear reversed evolution, i.e. the higher redshift, the higher density. This indicates an even stronger mass dependence in \GIZ; At outer radii, $r\gtrsim 0.8 \times r_{500}$, the similarity between \GIZ\ and \GX\ is shown up again with both decreasing with redshift. This indicates that the density profile is less sensitive to both the baryon models and halo masses at outer radii, which will not be suitable for constraining baryon models and characterizing cluster evolution. 

% In addition, the deviation in these different redshift comparisons becomes strong at the largest radius. Some cases even show a higher value than that at $z = 0$. This effect reflects that the matters at this far radius may have entered the domain of filaments which forms at an early time compared with clusters. 
% This effect reflects the relative amount of matter is dominated by the explosion of the universe at this far radius away from the cluster centre, i.e., more matter at higher redshift due to a smaller volume.

To clearly investigate the effects of baryon models on the density profiles, we further compare the profiles between \GX\ and \GIZ\ at the same redshift in the bottom panel of Fig.~\ref{fig:Gdens}. Consistently with the top panel, the largest difference can be $\sim 3$ times large and is mostly in the inner region. The difference between \GX\ and \GIZ\ seems smaller at higher redshift.
% except in the highest redshift bin which could be due to the resolution issue. 
The profiles in both simulations are basically consistent at $r \gtrsim 0.4r_{500}$ where the influence of central baryon models becomes weak. Note that as shown in \cite{Sorini2022} and \cite{Yang2023}, the gas particles ejected by the AGN winds in \GIZ\ are dumped much far away, $\gtrsim 2\times r_{200}$. That may be the reason why \GIZ\ shows little difference to \GX\ at out radii $r \sim r_{500}$. In addition, as the electron number density is more often used to describe the ICM distribution in observations than the gas density profile which requires additional assumptions, we further investigate the electron number density profile in the next section. 

%%%
% This suggests that the ICM component in the centre of galaxy clusters is more sensitive to the influences of sub-grid models.

% the only included thermal feedback models in \GIZ\ are less effective than both kinematic and thermal feedback contained in \GX.
% With redshift increasing, gas density begins to decrease especially since $z=3$ and deviate from self-similarity. Note that the lack of high-redshift profiles at the centre region is caused by the fewer gas particles.
% The ratios for \GIZ decrease from the inner region and begin to increase at $r\sim0.3 r_{500}$ all the time, while \GX shows increasing profiles at lower redshifts and become a descending tendency at $z > 3$. 
% Overall, the ratios increase with redshift decreasing, which is a result of halo growth. Meanwhile, the ratios in \GX\ ($z \leq 4$) and \GIZ\ ($z \leq 6$) respectively hold similar shapes evolution, which is an implication of self-similarity. This means that the self-similarity of gas mass density can hold at least $z = 4$.
% \qy{Besides, the ratios show a similar profile in both simulations at $0.5 r_{500} \lesssim r \lesssim r_{500}$, despite a difference in other radii due to the different times of mechanisms to trigger the baryon models.} At the outskirts of clusters, the ratio is steep and even large than 1 at $z < 3$. This corresponds to the high merger rate at this period.

\subsection{Electron number density} \label{sec:Eledens}
While electron number density is relatively easy to access from observations, we replenish a description for the distribution of the electron number density profile here. 
As the hot electrons in the ICM can scatter high-energy photons via bremsstrahlung emission or CMB photos through inverse Compton scattering, electron number density directly connects with X-ray observations and SZ effects, respectively. The electron number density is calculated using the following formula:  
\begin{equation}
    n_{\mathrm{e}} = \frac{1}{V m_{\rm p}} \sum_i (1 - Z_{\mathrm{He},i} - Z_{\mathrm{Metal},i})m_{\mathrm{gas},i}A_{\mathrm{e},i},
\end{equation}
where $Z_{\mathrm{He}}$ and $Z_{\mathrm{Metal}}$ are the abundance of helium and metallicity per gas particle, respectively. $A_{\rm e}$ is the number of electrons relative to the hydrogen atom per gas particle. $m_{\rm p}$ and $m_{\rm gas}$ are the mass of proton and gas, and $V$ represents the volume of shell at each bin.

\begin{figure}
    \centering
    \includegraphics[width=0.48\textwidth]{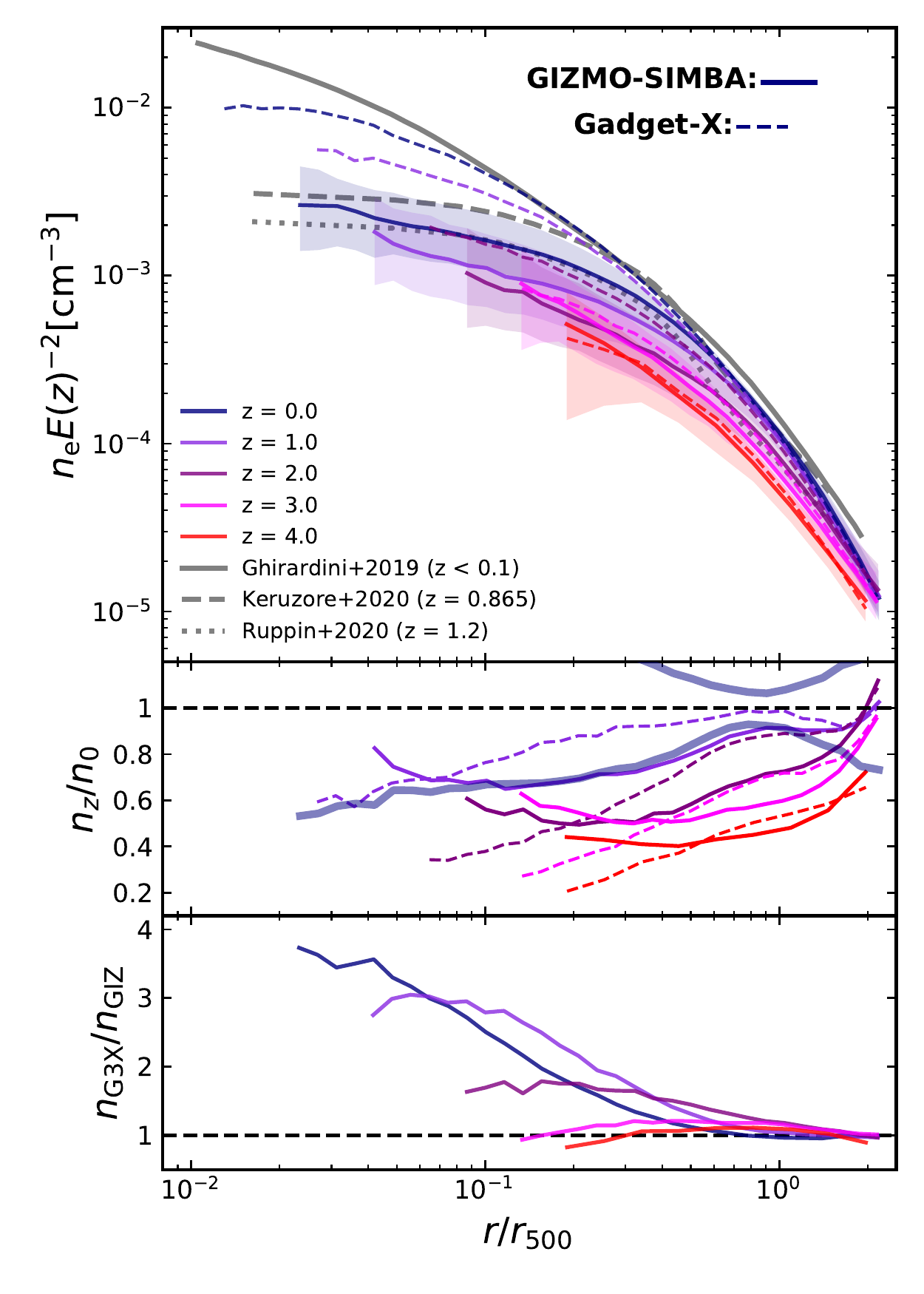}
    \caption{The evolution of electron number density scaled with $E(z)^{-2}$. The dotted, dashed and solid grey lines show the single cluster profile taken from \citet{Keruzore2020} ($z=0.865$), \citet{Ruppin2020b} ($z=1.2$) and \citet{Ghirardini2019} ($z<0.1$).}
    \label{fig:Eledens}
\end{figure}

In Fig.~\ref{fig:Eledens}, we show the electron number density rescaled with $E(z)^{-2}$, where $E(z)$ is redshift evolution factor defined as $E^2(z) = \Omega_m (1+z)^3 + \Omega_{\Lambda}$. We include the comparison with observations from \citet{Ghirardini2019} which analysed a sample of 12 low-redshift galaxy clusters selected from \emph{Planck} all-sky survey, and two single cluster profiles from SZ observation MOOJ1142+1527 \citep[$z$ = 1.2,][]{Ruppin2020b} and SZ/X-ray analysis of ACT-CLJ0215.4+0030 \citep[$z$ = 0.865,][]{Keruzore2020}. The hydrostatic mass of MOOJ1142+1527 is about $M_{500} \sim 6.1 \times 10^{14}\ \mathrm{M_{\odot}}$, while ACT-CLJ0215.4+0030 has a mass of $M_{500} \sim 3.5 \times 10^{14}\ \mathrm{M_{\odot}}$. Similar to Fig.~\ref{fig:Gdens}, both simulations are basically consistent with these observations at $r > 0.4 r_{500}$. It is expected that the electron number density profiles are highly similar to the gas density profiles, as the detected number of electrons is determined by the mass of ICM which is mostly hot gas. We checked the ratio between gas density and electron number density in both simulations. There is almost no evolution with the radius for \GX\ but a slight difference for \GIZ. This is because \GX\ lost the information of $Z_{\rm He}$ for each gas particle due to snapshot reproduction and the calculation for \GIZ\ is dependent on both $Z_{\rm He}$ and $Z_{\rm Metal}$. However, the fluctuation of the profile ratios in \GIZ\ is no more than 10 per cent in the entire radius. Therefore, our conclusions on the gas density profile from Fig.~\ref{fig:Gdens} also apply here.

\subsection{Temperature} \label{sec:temp}
\begin{figure}
    \centering
    \includegraphics[width=0.48\textwidth]{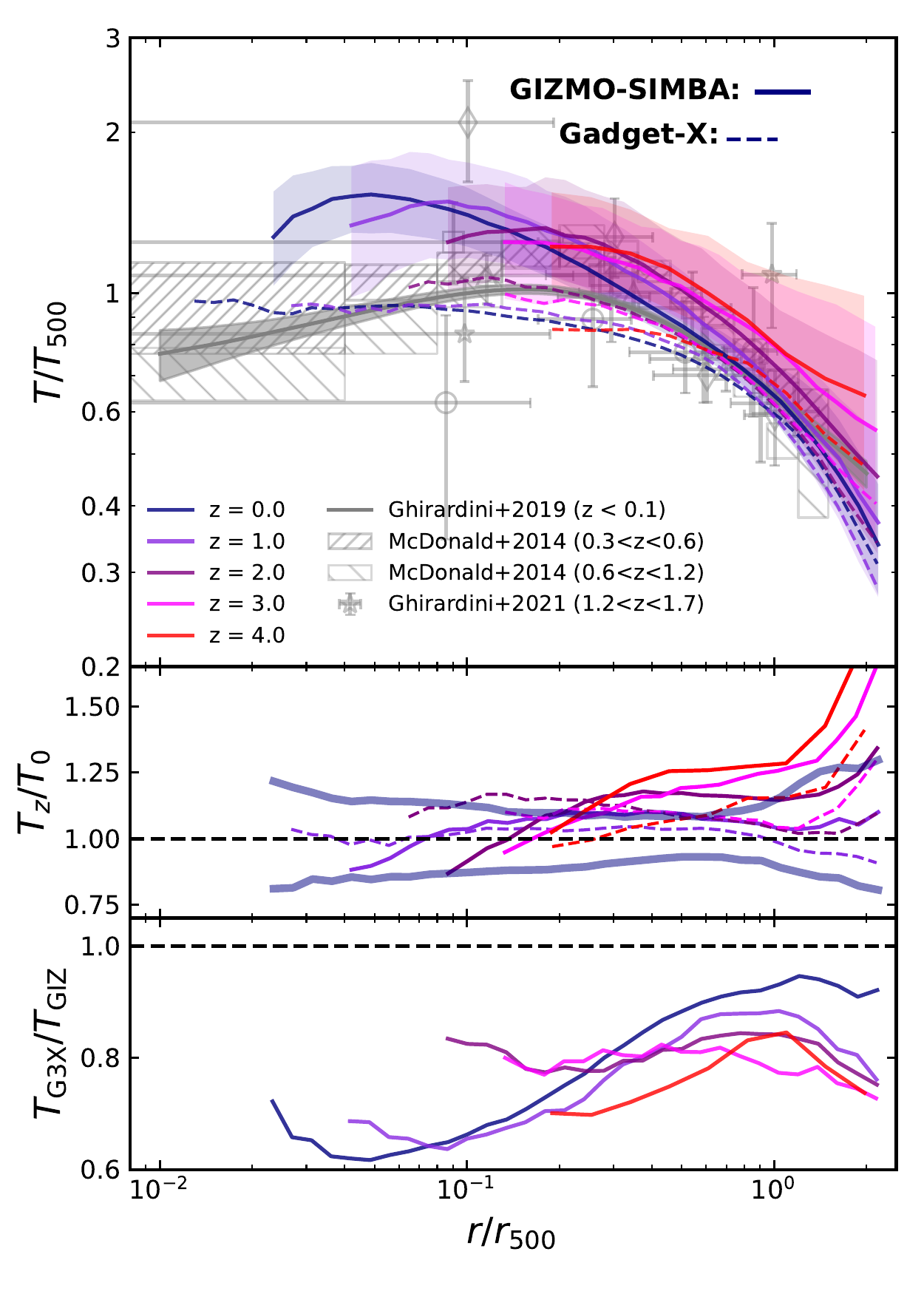}
    \caption{The evolution of mass-weighted temperature scaled with $T_{500}$. The error bars with different hollow symbols represent eight individual cluster profiles from \citet{Ghirardini2021} at a high redshift bin. The rectangles filled with a slash are the observations at two redshift bins $0.3<z<0.6$ and $0.6<z<1.2$ from \citet{McDonald2014}. The solid grey line with the shaded region shows the results taken from \citet{Ghirardini2019} at $z<0.1$.}
    \label{fig:MWTemp}
\end{figure}

In this section, we focus on temperature profiles which are popularly included to derive cluster mass through hydrodynamic equilibrium equation \citep[e.g.,][]{Rasia2004} or scaling relation \citep[e.g.,][]{Cui2018}.

In Fig.~\ref{fig:MWTemp}, we present the median mass-weighted temperature profiles scaled with $T_{500}$. The definition of $T_{500}$ from \citet{Ghirardini2019} is expressed as:
\begin{equation}
    T_{500} = 8.85\ \mathrm{keV} \left(\frac{M_{500}}{10^{15}\ h^{-1}_{70}\mathrm{M_\odot}}\right)^{2/3} E(z)^{2/3} \frac{\mu}{0.6},
    \label{T500G}
\end{equation}  
where $\mu$ is the mean molecular weight per gas particle, which \citet{Ghirardini2019} assumed to be equal to 0.6125 \citep{Anders1989}. The simulated profiles are compared with observational results, including \citet{Ghirardini2019} which presented the thermodynamic property of nearby ICM with the X-COP sample; \citet{McDonald2014} which showed the redshift evolution of mean temperature with 80 X-ray and SZ joint galaxy clusters at two redshift bins $0.3 < z < 0.6$ and $0.6 < z < 1.2$; \citet{Ghirardini2021} which calculated the thermodynamic properties of 7 high-redshift galaxy clusters with a total SZ-inferred mass $M_{500} > 3 \times 10^{14}\ \rm M_{\odot}$ out to $z \sim 1.8$. The cluster sample in \citet{McDonald2014} has a mean $M_{500}$ of $\rm 5.5 \times 10^{14}\ M_{\odot}$ and $\rm 4.2 \times 10^{14}\ M_{\odot}$ in the low and high redshift bin, respectively. \citet{McDonald2014} and \citet{Ghirardini2019} also studied the entropy and pressure profiles with the same cluster data, which are shown in our investigation on entropy and pressure profiles in the following sections.

The simulated temperature profiles from \GIZ\ and \GX\ at low redshifts are also similar at a radius $r \gtrsim 0.4r_{500}$ (with a difference of about 20 per cent). In the core region, $r \lesssim 0.1r_{500}$, \GX is in better agreement with the observation data albeit with a slightly higher temperature at these most inner data points. In contrast, \GIZ tends to be closer to observation data at intermediate radii $0.2r_{500} \lesssim r \lesssim 0.4r_{500}$. However, as redshift increases, \GIZ\ presents an increasing temperature profile, while \GX\ presents an almost unchanged tendency. This in-conformity of temperature at high redshifts may be caused by the different influences of baryon models, which needs to be confirmed with deep observations. As shown in Fig.~\ref{fig:MWTemp}, the error bars for these observation data are still too large and the variations between different clusters also make it hard to distinguish the small evolution.
In the core, \GX\ shows a lower and flatter temperature profile compared with \GIZ. We consider this difference due to the AGN feedback models adopted in \GIZ, i.e., gas particles are heated up by AGN outflow, which is consistent with the performance of ICM mass density (Fig.~\ref{fig:Gdens}). Meanwhile, if more gas is blown away, the remaining gas heats up as it flows inwards to maintain hydrostatic equilibrium. It is worth noting that the temperature profile, however, peaks at too inner radius compared to the observational results, which indicates that the AGN feedback in \GIZ may be dominated by the radiative mode instead of jet mode due to too much hot gas surrounding the central BH, resulting in a high accretion rate. Further, the temperature in a simulated cluster at $z = 0.3-0.5$ from \citet{Tremmel2019}, which also imparted kinetic term in AGN feedback, presented a similar temperature shape as \GIZ\ shown.

In the middle panel of Fig.~\ref{fig:MWTemp}, we can find that the self-similarity of temperature profiles in \GX\ holds at all redshift ranges and extends within $0.1r_{500}$, while the profiles in \GIZ\ begin to deviate from self-similarity at $z \sim 2$. We use the ratio between the percentile errors and median value of density at $z = 0$ as the criterion for judging self-similarity as shown with bold navy blue lines. Besides, the redshift evolution for \GX\ seems always within the errorbar region at $z=0$. Again, a clear redshift evolution of the gas temperature profile is shown in \GIZ\ than \GX. 

The profile differences between the two simulations are shown in the bottom panel of Fig.~\ref{fig:MWTemp}. We see a higher temperature profile in \GIZ compared with that in \GX\ over the total radius range with an even larger difference at higher redshift. This seems a contradiction to the result in Fig.~\ref{fig:fgas_evo} at $z=0$, i.e. more hot gas in \GX than in \GIZ. The discrepancy should be (1) most of the hot gas in \GX\ has actually a lower temperature than in \GIZ \citep[see Figure A1 in][]{Cui2022}; (2) caused by the normalization of $T_{500}$, which is proportional to $M_{500}$. As shown in \cite{Cui2022EPJWC}, \GX has a slightly higher $M_{500}$ than \GIZ. Furthermore, as the gas temperature is more sensitive to the implemented feedback models than gas density \citep{Truong2021, Yang2023}, the gas temperature profile at high redshift would be useful for future constrain of the baryon models. Though we need to theoretically understand which model is mostly responsible for this difference, this requires a lot of work, i.e. turning on and off each baryon model to isolate its effect, which will be studied in another paper. 

In comparison to the fixed halo mass evolution (see Fig.~\ref{fig:temp_appd}) which shows a reversed trend for both simulations, we think the higher normalized temperature profile at high redshift in Fig.~\ref{fig:MWTemp} should be caused by the normalization $T_{500}$. In other words, the physical gas temperature in clusters with the same mass at high redshift should be lower as they still need time to be thermalized.

\subsection{Entropy}

\begin{figure}
    \centering
    \includegraphics[width=0.48\textwidth]{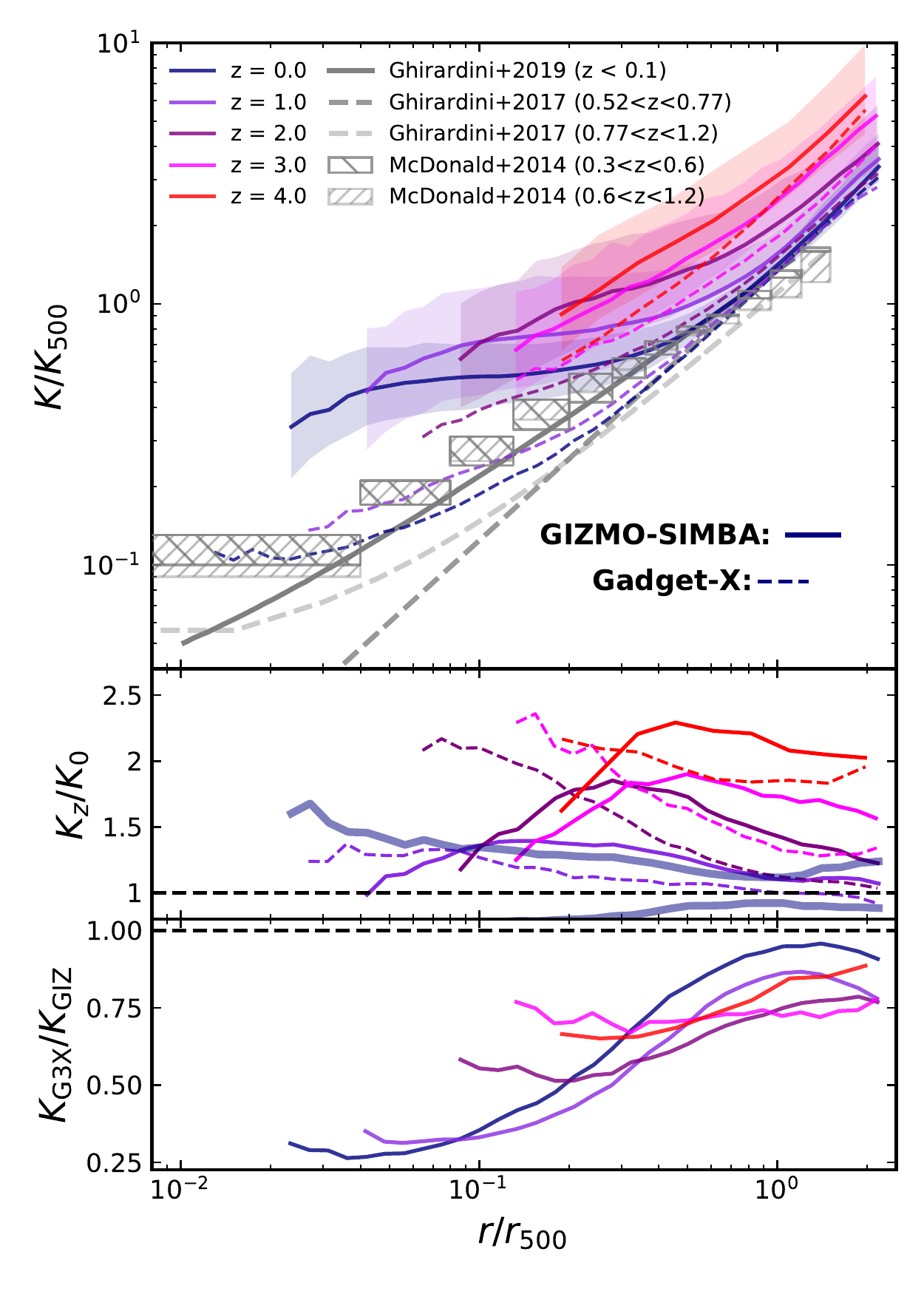}
    \caption{The evolution of entropy profiles scaled with $K_{500}$. The rectangles filled with a slash are the observations at two redshift bins $0.3<z<0.6$ and $0.6<z<1.2$ from \citet{McDonald2014}. The solid grey line shows the results from \citet{Ghirardini2019} at $z<0.1$. The deep and shallow dashed grey lines represent the results in \citet{Ghirardini2017} at $0.52<z<0.77$ and $0.77<z<1.2$ respectively.}
    \label{fig:Entropy}
\end{figure}

The distribution of ICM entropy provides a unique view to exploring the thermal history of ICM. While the primordial gas is isotropic according to the standard model of structure formation, the subsequent distribution of entropy carries on the evolution information of ICM after experiencing disturbance, such as the radiative and dissipative processes, other non-gravitational sources or sinks of energy and the shock waves \citep[e.g.][]{Finoguenov2007, Planelles2014}.

We calculate the entropy with the following definition:
\begin{equation} \label{eq:entropy}
    K = \frac{k_{\mathrm{B}} T}{n_\mathrm{e}^{2/3}},
\end{equation}
where $k_{\rm B}$ is the Boltzmann constant. 
In Fig.~\ref{fig:Entropy}, we present the entropy profiles by scaling to $K_{500}$ \citep{Ghirardini2017}, which is defined as:
\begin{equation}
    K_{500} = 103.4\ \mathrm{keV\ cm^2} \left(\frac{M_{500}}{10^{14} \rm M_\odot}\right)^{2/3} E(z)^{-2/3}f_{\rm b}^{-2/3},
    \label{K500}
\end{equation}  
where the baryonic fraction $f_{\rm b}$ is adopted as 0.15 \citep{Planck2016}. In addition to the observations from \citet{McDonald2014} and \citet{Ghirardini2019}, we also include the comparison to \citet{Ghirardini2017} which calculated the entropy and pressure profiles with 47 galaxy clusters selected from \Chandra\ X-ray observations at $0.4 < z < 1.2$. The $M_{500}$ in their sample is larger than $10^{14}\ \rm M_{\odot}$. Note that the observational results from \cite{Sun2009, Pratt2010} are not included due to the relatively smaller cluster masses of their samples.

The entropy in both simulations is steep in the outskirts, distributed in a power-law profile as expected. The profiles from both simulations and observations are consistent with each other at low redshifts. The self-similarity in both hydro simulations holds to $z\sim 2$ at $r \gtrsim 0.6r_{500}$, which can be verified from the entropy ratio in the middle panel of Fig.~\ref{fig:Entropy}. As redshift increases, entropy shifts up at a similar rate for both \GIZ\ and \GX. 
% besides the highest redshift $z=6$. 
This indicates that the electron number density rather than temperature is more responsible for the redshift evolution. This can be further viewed in the bottom panel of Fig.~\ref{fig:Entropy} as the difference between the two simulations is almost independent on redshift.

In the core region, $r\lesssim 0.2r_{500}$, \GX\ tends to follow the power law slope which is generally consistent with low-redshift observations, while \GIZ\ shows higher and flatter profiles. This higher value also called an excess to the self-similar prediction, is observed in the observations for non-cool-core clusters \citep[e.g.,][]{Tozzi2001,Bulbul2016,Walker2019,Ghirardini2021}, which is thought to be caused by non-gravitational processes. Nevertheless, the bimodality separation in cool-core and non-cool-core clusters has been identified in \GX (note that no error bars have been shown for \GX here), while it seems that it is unlikely to find cool-core clusters in \GIZ.
This entropy core problem has also been shown in recent hydrodynamical simulations such as C-EAGLE \citep{Barnes2017b}, TNG \citep{Barnes2018}, MTNG \citep{Pakmor2022} and FABLE \citep{Henden2018}, despite which types of AGN models (thermal, kinetic, or mixed thermal and kinetic) are considered. While the clusters in MACSIS \citep{Barnes2017a}, which implemented AGN feedback with a purely thermal scheme, produced a power-law entropy profile at low redshift. Besides, some works without metal cooling can also generate a low entropy core \citep{Tremmel2019, Altamura2022}. However, metal cooling is not a necessary mechanism to overcome the core problem, as \GX\ considers metal cooling but still shows a power-law entropy profile. Using a new implementation of the EAGLE galaxy formation model, \cite{Altamura2022} claimed that this problem seems to be unrelated to incorrect physical assumptions, missing physical processes or insufficient numerical resolution. Note they adopted a purely thermal AGN feedback implementation in all simulations. We suspect this problem may be caused by the coupling between baryon models and resolution or is a result of multi-non-gravitational effects which is hard to be solved by only controlling one variable. While we find this problem is more resolution-dependent (Cui et al. in prep.).
% \cite{Oppenheimer2021} for galaxy groups with the explanation that could be its decoupled kinetic AGN feedback scheme.   
At last, the excess increases with redshifts in both simulations and observations, which is much weaker for observation and \GX\ within $z=1$.

A similar conclusion as the fixed halo mass temperature evolution, can be drawn for the fixed halo mass entropy evolution (see Fig.~\ref{fig:entropy_appd}). Given that the gas temperature decreases and density increases with redshift, it is not surprising to see the entropy profile decreases with increasing redshift.

\subsection{Pressure}
\begin{figure}
    \centering
    \includegraphics[width=0.48\textwidth]{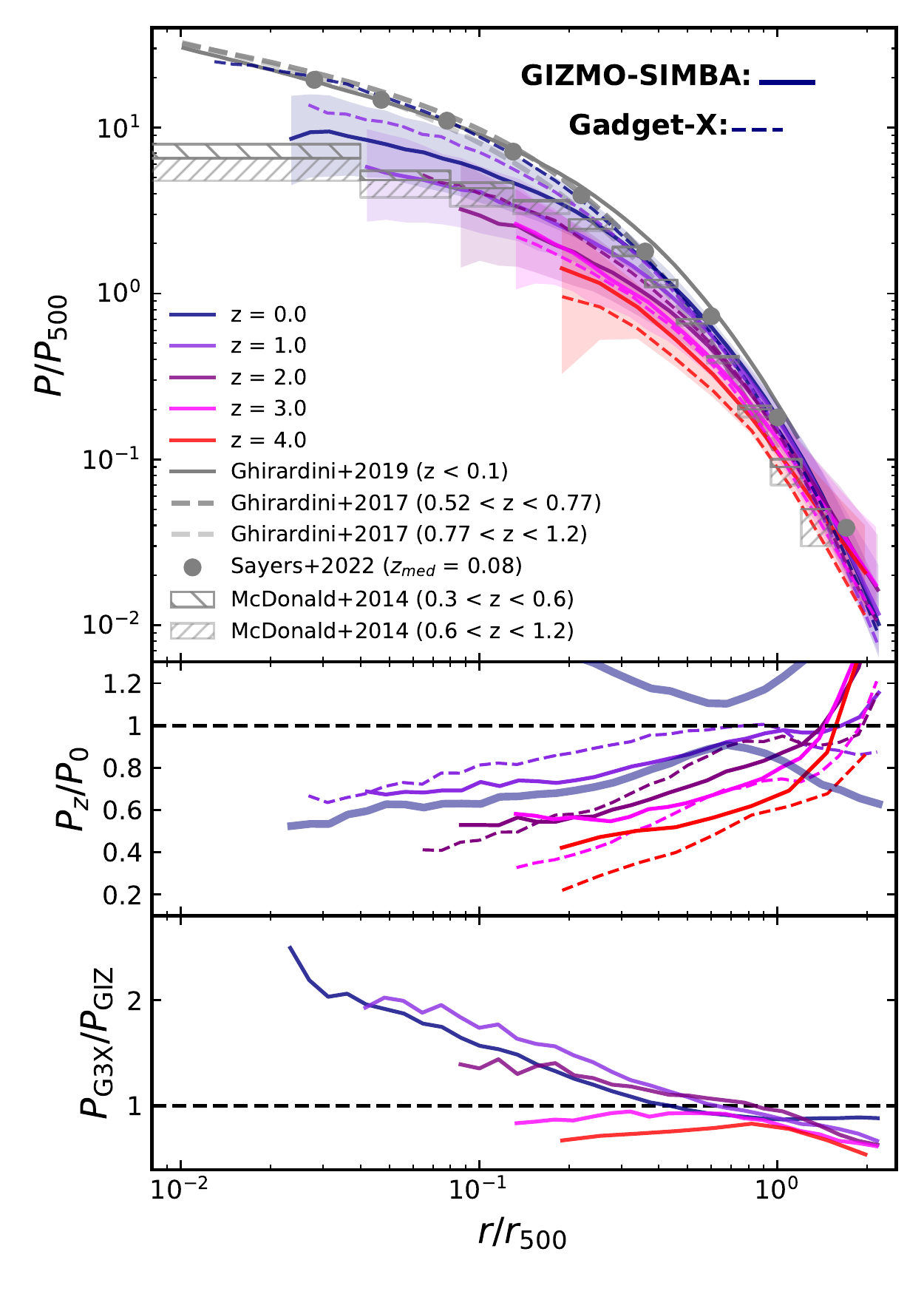}
    \caption{The evolution of pressure profiles scaled with $P_{500}$. Note the observations shown with grey colours use the same data as shown for the entropy in Fig.~\ref{fig:Entropy}. Comparisons to the observational results at different redshifts: \citet{McDonald2014,Ghirardini2017,Ghirardini2019,Sayers2023} are shown in grey regions, lines and symbols as indicated in the legend. }
    \label{fig:Pressure}
\end{figure}

As an essential indicator of non-gravitational processes to compete with the gravitational force, pressure is directly connected to the depth of the gravitational potential \citep{Arnaud2010}. Thus, the distribution of ICM pressure can reflect the competing results between the gravitational and non-gravitational processes. The gas pressure profile can be measured through the SZ signal and also be used to make predictions for the cluster mass \citep[see][for example]{Gianfagna2021} under the hydrostatic assumption.

We calculate pressure according to the formula provided from \citet{Planelles2017},
\begin{equation}
    P = \frac{k_{\mathrm{B}}}{\mu m_{\mathrm{p}} V}\sum_i m_{\mathrm{gas}, i} T_i.
\end{equation}
In Fig.~\ref{fig:Pressure}, we present the pressure profiles, which is scaled to $P_{500}$ defined in \citet{Nagai2007},
\begin{equation}
    P_{500} = 1.45 \times 10^{-11} \mathrm{erg\ cm^{-3}} \left (\frac{M_{500}}{ 10^{15}h^{-1} \rm M_{\odot}} \right)^{2/3} E(z)^{8/3}.
\end{equation}
The comparisons to observations adopt the same data as shown for entropy in Fig.~\ref{fig:Entropy}.

The radial pressure distribution is similar to the density profiles (Fig.~\ref{fig:Gdens}), showing a high consistency in simulations and observations.
As pressure is calculated based on the density and temperature, thus the shape of pressure is expected to follow the combination of density and temperature profiles. 
In addition, since gas density is the result of the competition between gravitational and non-gravitational processes, and the pressure mainly reflects gravitational potential, we expect to obtain the clues of complex baryonic processes by comparing density and pressure profiles. 
The slope of the pressure profile is shallow in the centre and becomes steep in the outskirts. The scatter gradually becomes small with the radius increasing. At the core region, \GX matches well with \cite{Ghirardini2019}, as well as \cite{Sayers2023} at $z\sim0$, while \GIZ tends to agree with \cite{McDonald2014}. Both are consistent with observations at the outskirts. As also indicated in the bottom panel of Fig.~\ref{fig:Pressure}, there seem smaller differences between \GIZ and \GX compared to their density profile changes. This could be because of the slightly increased gas temperature in \GIZ, which can compensate for their density decrease. Furthermore, the pressure profile reflects the potential of the halo. Thus, it should also be less sensitive to the details of the subgrid models for these astrophysical processes in gas evolution. As shown in the middle panel of Fig.~\ref{fig:Pressure}, the self-similarity is obvious beyond $r_{500}$ even at $z \sim 3$. While this outer region is less affected by the non-gravitational process, thus self-similarity is an expression of the gravitational process. The two simulations are generally in agreement at large radii in different redshifts. 

Interestingly, the fixed halo mass evolution for pressure profile (see Fig.~\ref{fig:pres_appd}) agrees with Fig.~\ref{fig:Pressure}, i.e. higher redshift tends to present a lower pressure albeit a much clear separation between different redshifts even at $\sim r_{500}$. This indicates a pressure increment along with their thermalization.

% \citep{Voit2005},
% \begin{equation}
% \begin{split}
%     P_{500} = 3.426 \times 10^{-3} \mathrm {keV\ cm^{-3}} \left (\frac{M_{500}}{h_{70}^{-1} 10^{15} M_{\odot}} \right)^{2/3} E(z)^{8/3} \\
%     \times \left(\frac{f_b}{0.16}\right) \left(\frac{\mu}{0.6}\right) \left(\frac{\mu_e}{1.14}\right),
% \end{split}
% \end{equation}
% where the mean molecular weight $\mu$ is 0.5994, the mean molecular mass per electron $\mu_e$ is 1.1548 and $f_b$ is the baryon fraction.

\subsection{Metallicity} \label{sec:metal}
\begin{figure}
    \centering
    \includegraphics[width=0.48\textwidth]{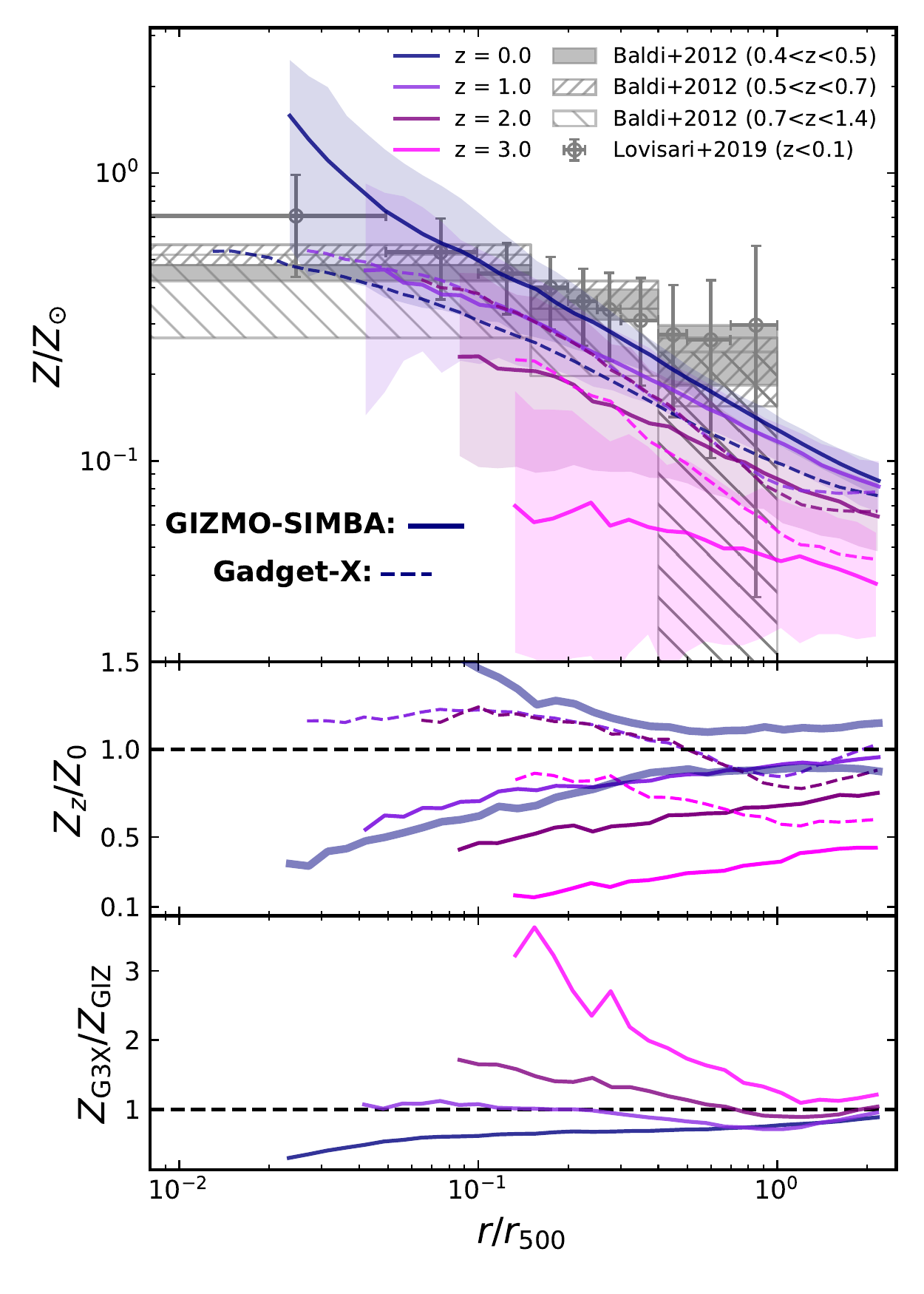}
    \caption{The evolution of mass-weighted metallicity. The error bars with grey hollow circles show the median abundance profile and scatter of measurements taken from \citet{Lovisari2019}. The shaded regions filled with different slashes represent the average abundance obtained from \citet{Baldi2012} at three different redshift bins.}
    \label{fig:Metal}
\end{figure}

Metallicity is a useful quantity to trace the galaxy formation history with the cumulative chemical enrichment produced from supernova feedback and stellar nucleosynthesis, which makes a contribution to cluster formation and evolution.   

In Fig.~\ref{fig:Metal}, we show the median mass-weighted gas metallicity profiles from $z=3$ to $z=0$. The metallicity is normalized to the solar metallicity, $Z_\odot = 0.0134$ \citep{Asplund2009}. The low-redshift metallicity profiles are compared with the stacked profile in \citet{Lovisari2019}, who derived the metallicity profiles for 207 galaxy groups and clusters at $z<0.1$ with the observation data from \Newton. We also include the comparison with \citet{Baldi2012}, who presented the evolution of spatially-resolved metal abundance with 39 X-ray spectra galaxy clusters from \Newton\ at $0.4<z<1.4$. Here we select three redshift bins ($0.4<z<0.5$, $0.5<z<0.7$ and $0.7<z<1.4$) shown with different shadow patterns.  

At $z=0$, \GIZ has a higher metallicity profile compared to \GX at all radii. This is also consistent with the findings from stellar metallicity profiles (see Section.~\ref{sec:stellar_metal}) albeit with a slightly larger difference. \GIZ also exhibits a much higher metallicity core within $r \lesssim 0.1r_{500}$ compared to observational results, which agree with \GX better. Note that \cite{Ettori2015} reported that cool-core clusters tend to have a higher metallicity compared to non-cool-core clusters at the cluster centre. Though it is a little confusing here that \GIZ\ can still have a high metallicity core with almost all the clusters being non-cool-core clusters, we suspect this is because \GIZ\ has a higher star formation at high redshifts, thus more metals are released from stars.
At $r \gtrsim 0.1 r_{500}$, the metallicity profiles from both simulations decline with radius; \GIZ tends to have a better agreement to observational results than \GX, however, observed metallicity profiles are much flatter than the simulated ones, especially at outer radii. 

There is almost no redshift evolution for \GX out to $z \approx 2$, which is in agreement with other simulations \citep[e.g.,][]{Vogelsberger2018, Biffi2018, Pearce2021}, though a little lower metallicity profile at $z = 3$. However, \GIZ presents a clear and gradual redshift dependence with lower metallicity profiles at higher redshift. As the metal enrichment history is tightly connected with star formation and SN feedback, this difference suggests that the metallicity profiles can be used to pin down the baryon models. However, with the large errorbar in observation data, it is still not clear whether there is a redshift evolution or not\footnote{Though observational results preferred a no redshift evolution out to $z\sim 1$, for example, \cite{McDonald2016,Mantz2017}, it is worth noting that the metallicity profile depends on the cluster dynamical state and galaxy cluster and group have different metallicity profiles \citep{Lovisari2019}, which makes it hard to draw a solid conclusion.}. Though the iron (Fe) abundance is not resolved in \GX, the radial distribution of Fe abundance in \GIZ\ generally follows the shape of metallicity profile with a lower value, which is consistent with the results in other simulations \citep[e.g.,][]{Vogelsberger2018, Pearce2021}. Also note here that we used mass-weighted metallicity, whereas observations usually adopt emission-weighted metallicity, we refer interesting readers to \cite{Biffi2018}, which concluded: "A flatter shape towards the outskirts is also recovered, when the emission-weighted metallicities are computed". Furthermore, \citet{Biffi2018a} studied projected emission-line weighted and 3D mass-weighted metallicity and found a lower Fe metallicity and decreased scatter at the outskirts of the cluster in mass-weighted profiles. The difference $\sim 0.1\ \rm Z_{\odot}$ suggests that a projected emission-line weighted profile should be used for a proper comparison with the observational result. However, this requires proper mock X-ray observation maps, which are not included in this theoretical study for we are mostly interested in the differences between \GIZ\ and \GX. Furthermore, the chemical evolution model in \GIZ\ is still not perfect, individual metal evolution and abundance will be investigated in the following paper with the high-resolution clusters simulated with the latest version -- \textsc{SIMBA-C} (Hough et al. in prep.).

It is not surprising to see the metal enrichment in ICM through tracking. Even the metallicity profiles with fixed halo mass in Fig.~\ref{fig:metal_appd} indicate that there are fewer metals in ICM at higher redshift. It is unclear whether there is a halo mass-dependent or not on the metallicity profile from these two plots. We verified that there is a slight halo mass-dependent by comparing the cluster density profiles from different mass haloes at $z=0$.

\section{Stellar profiles} \label{sec:stellar}

\begin{figure}
    \centering
    \includegraphics[width=0.48\textwidth]{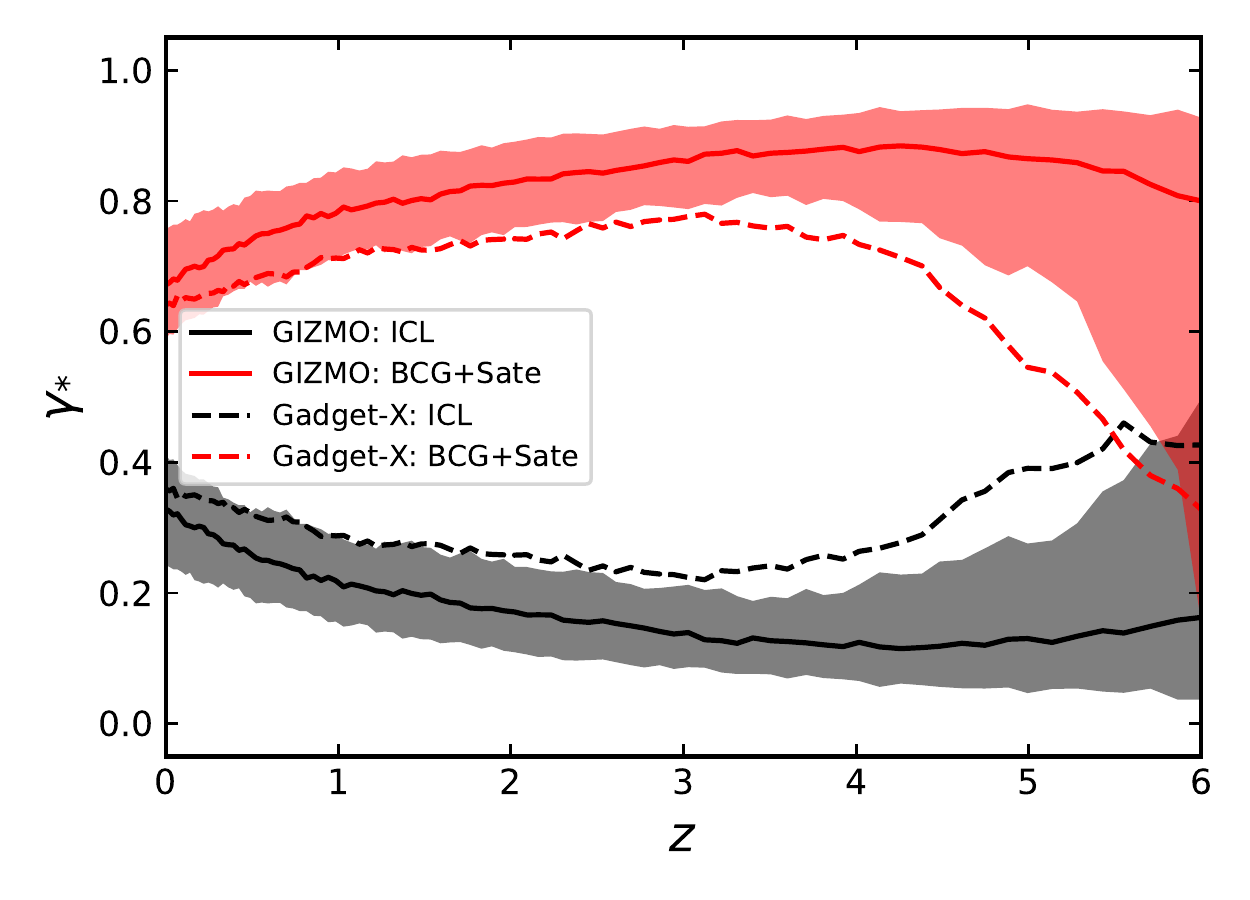}
    \caption{The abundances of the stellar mass in different environments as a function of redshift. The solid and dashed lines indicate the evolution in \GIZ\ and \GX, while the black and red lines show the stellar particles in BCG and satellite galaxies and ICL, respectively. The shaded regions are the 16th and 84th percentiles of all clusters in \GIZ.}
    \label{fig:fsubste_evo}
\end{figure}

Although the stellar component only occupies about a few per cent of the total cluster mass, they are another major observable for us to investigate galaxy clusters. In this section, we present a theoretical study of the evolution of the stellar distribution in galaxy clusters. 

First, we explore the ratio of a stellar mass fraction under different bound conditions as shown in Fig.~\ref{fig:fsubste_evo}. Here, the definition of mass fraction ratio is similar to gas (Equation ~\ref{eq:gamma}) but for stellar component, i.e., $\gamma_* = M_X/M_*$, where $M_*$ is the total stellar mass within $r_{500}$. The definition of BCG and satellite galaxies is the same as that we use to study gas distribution (Section.~\ref{sec:gas_history}), while the star particles in the diffuse component are regarded as intracluster light \citep[ICL, see][for example]{Cui2014}. Overall, the stellar mass in BCG and satellite galaxies is always higher than that in ICL. There is even a higher stellar mass fraction in galaxies in \GIZ than that in \GX. As there is a similar amount of stellar mass between \GIZ and \GX at $z=0$ \citep[see Fig. 2 in][]{Cui2022}, \GX seems to assign more stellar mass to ICL than \GIZ, or more stars in galaxies and within $0.05 r_{500}$ in \GIZ. Though we consider that \GIZ\ has a strong AGN feedback, this higher $\gamma_*$ in galaxies indicates that the stellar distribution outside the galaxy range may be unaffected by this. Note that it is not clear whether the BCG definition ($<0.05 r_{500}$) at high redshift is suitable or not. The high ICL fraction for \GX at $z =6$ could be caused by this definition and/or a much lower stellar mass at high redshift compared to \GIZ \citep[see Fig. 2 in][]{Cui2022}. In addition, the different fractions in ICL also relate to details of the star-formation and feedback models in the two simulations. However, this investigation is only to get a rough idea of how star particles are distributed. More detailed investigations on the ICL in the300 clusters are in final preparation (Contreras-Santos, et al.). In summary, the two simulations show a consistent evolution of the global quantities, but with a slightly different assignment in stars, especially at high redshift. In the following, we focus on three detailed distributions of stellar properties: the stellar mass density, the satellite number density and the stellar metallicity.

\subsection{Stellar mass density}

\begin{figure}
    \centering
    \includegraphics[width=0.48\textwidth]{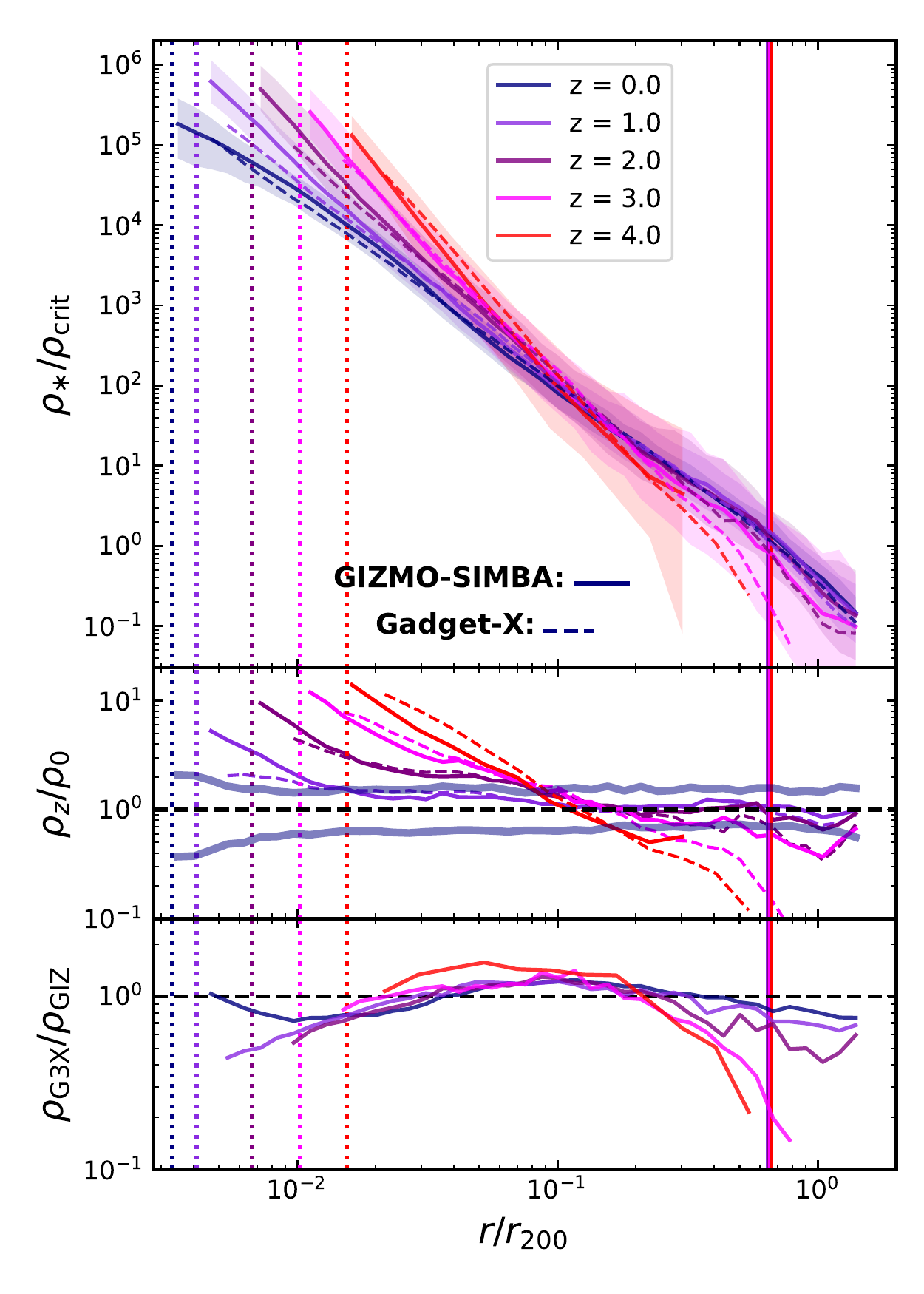}
    \caption{The evolution of stellar mass density scaled with the critical density $\rho_{\rm crit}(z)$ of the universe and the radius scaled to $r_{200}$. The vertical solid and dotted lines respectively show the median value of $r_{500}/r_{200}$ and softening length in \GIZ with the colours same as the shown profile at the corresponding redshift. Profiles are truncated approaching the softening length. The value of $r_{500}/r_{200}$ at different redshifts is very close to each other. Note the median value of $r_{500}/r_{200}$ in \GX\ is almost identical to that in \GIZ.}
    \label{fig:Stedens}
\end{figure}

The stellar mass density profile further details how all the stars within galaxies and ICL are distributed inside galaxy clusters. In Fig.~\ref{fig:Stedens}, we present the median stellar mass profiles from $z=4$ to $z=0$. The stellar mass density is scaled to the critical density of the universe. Note we here normalize the radius to $r_{200}$ instead of $r_{500}$, which is visible by the vertical solid lines ($\sim 0.65 r_{200}$). The vertical dotted lines mark the positions of softening length for \GIZ at different redshifts. Note that the positions of softening length for \GX\ are slightly larger than that shown in \GIZ, which have been detailed in \autoref{sec:data}, however, are not included in this figure for simplicity.  
% Also, note that we show a slightly larger radius range due to that $r_{200}$ is used.
% in order to be a convenient of comparison with observations.
At the four redshifts, \GIZ\ seems to have a higher stellar mass density than \GX in the cluster centre $\lesssim 0.02 r_{200}$. This can be explained by an early time of consuming more gas into stars in \GIZ, which will form a dense core from the very beginning. This slightly higher stellar density can be kept even after a strong feedback model quenches the galaxy in \GIZ. 
The picture is also supported by the large gap at the very centre $\lesssim 0.01 r_{200}$ in the stellar density between $z = 0$ and at $z = 1$ during which the BCG in \GIZ probably gets quenched, while \GX\ has similar values. This also implies that the efficiency of star-forming in the centre is quite different and sensitive to the baryon models adopted in the two hydro runs. 
Besides, the deviation from the profile at $z = 0$ as shown in the middle panel of Fig.~\ref{fig:Stedens} are mostly contributed by mass evolution through a comparison with Fig.~\ref{fig:stedens_appd}.
% The gaps (i.e., strong evolution of stellar density) in \GIZ\ at different redshifts could also be the consequence of the strong feedback. 

When the radius becomes larger, up to $\sim 0.2 r_{200}$, the differences between the two simulations and among different redshifts seem smaller compared to the central region, which is highlighted in the bottom panel. Furthermore, it is interesting to note that the slope of the stellar profile becomes steep from low to high redshift. This result is in agreement with the observation result in \cite{van2015}, which showed a steeper slope at $z=1$ compared to their $z=0$ stellar mass density profile. Furthermore, after normalized to $z=0$ profile in the middle panel of Fig.~\ref{fig:Stedens}, these changes seem to rotate around an axis at $\sim 0.1r_{200}$, i.e., a highly constrained point despite different redshifts and baryon models. This suggests that the star formation is less affected by the gravitational and non-gravitational processes at this radius. Note that the crossing point is also very similar (about 300 kpc if we assume $r_{200} = 2\ h^{-1}\rm Mpc$) to the one shown in Fig. 7 in \cite{van2015}.

\subsection{Satellite galaxy number density}

\begin{figure}
    \centering
    \includegraphics[width=0.48\textwidth]{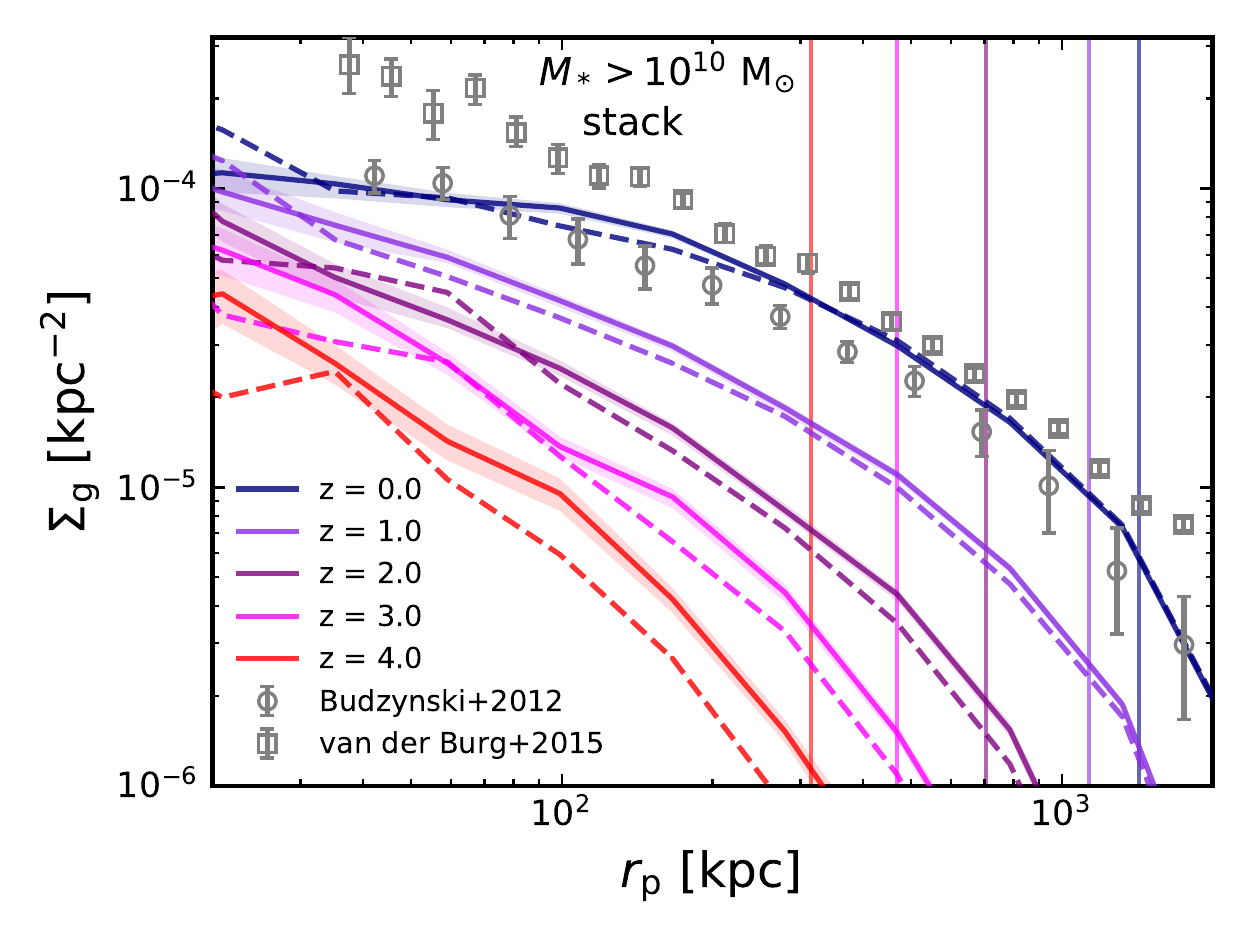}
    \caption{Projected satellite number density profiles. The solid and dashed lines respectively show the distribution of galaxies in \GIZ\ and \GX with a stellar mass larger than $ 10^{10}\ \rm M_{\odot}$, while the shaded region shows the 16th and 84th percentiles from 100 bootstrapping samples. The number density is calculated from a stacked sample. The error bars with grey circles and squares present the observations taken from \citet{Budzynski2012} ($0.15 \leq z < 0.4$) and \citet{van2015} ($0.04 < z < 0.26$), respectively. The vertical lines mark the median $r_{500}$ for the cluster sample in \GIZ at each redshift, with the colours the same as the shown profile at the corresponding redshift.}
    \label{fig:Ngal_prop}
\end{figure}

The previous section only presents a theoretical study of the stellar distribution in 3D. However, it is impossible to directly compare the simulated stellar distribution with observational results from galaxies. Satellite galaxy number density profile, instead of mass density profile, allows easy comparison with observation but also reveals cluster properties, such as cluster dynamical state \citep[e.g.][]{DeLuca2021} or merger history, from a different point of view. Therefore, we explore the evolution of the number density of satellite galaxies along with the formation of the cluster here. 

In order to compare with observational results, we project a spherical region with a radius of $1.5r_{200}$ at the centre of galaxy clusters, along the z-axis for this case. Only satellite galaxies with mass $M_* > 10^{10}\ \rm M_{\odot}$, a similar stellar mass cut for these observational results shown in Fig.~\ref{fig:Ngal_prop}, are used. Since we stack satellite galaxies from each cluster due to their small number, especially at high redshift, all satellite galaxies are counted in a logarithm bin without any clusters discarded. The total number of satellite galaxies in each bin is divided by the projected area and the total number of clusters to obtain satellite number density, $\Sigma_\mathrm{g}$. Note that the comoving distance is used here. We bootstrap our cluster sample 100 times to produce a sample of newly stacked satellites. These samples are then used to calculate 16th and 84th-percentile scatters. We include comparisons to low-redshift observations from \citet{Budzynski2012} (B12) and \citet{van2015} (vdB15) in Fig.~\ref{fig:Ngal_prop}. B12 used a group and cluster catalogue from SDSS 7 in the redshift $ 0.15 \le z \le 0.4$. The halo mass range in this compared sample is $10^{14.7}\ \mathrm{M_{\odot}} < M_{500} < 10^{15}\ \rm \mathrm{M_{\odot}}$ approaching to ours at $z=0$ as shown in Table.~\ref{tab:1}. Their satellite galaxies are selected with a modest magnitude limit $M_r = -20.5$. To make a comparable contrast, we estimate the corresponding stellar mass criterion based on the stellar mass--absolute magnitude relation in \citet{Mahajan2018}, which derived the relation using a low-redshift galaxy sample from the Galaxy and Mass Assembly survey (GAMA) at $r$ band. The formula to derive galaxy stellar mass from absolute $r$-band magnitude is expressed as $\log M_*/\mathrm{M_{\odot}} = 0.450 - 0.464M_r \pm 0.458$. Based on this formula without an error consideration, we obtain the stellar mass limit in B12 corresponding to $9.16 \times 10^{9}\ \rm M_{\odot}$, a slightly lower value than the one predicted in vdB15, which included 60 massive clusters in $10^{14}\ \mathrm{M_\odot}  \gtrsim M_{200} \gtrsim 2\times 10^{15}\ \mathrm{M_\odot} $ with a mean value of $\sim 7 \times 10^{14}\ \mathrm{M_\odot}$ at 0.04 < $z$ < 0.26 selected from the Multi-Epoch Nearby Cluster Survey (MENeaCS) and the Canadian Cluster Comparison Project. vdB15 used all the galaxies with $M_* > 10^{10}\  \rm M_\odot$, the same as we adopted for the simulated galaxies. However, to compare with the data in Fig.~\ref{fig:Ngal_prop} consistently, we need to remove their normalization of $r_{200}^{-2}$ on $\Sigma_\mathrm{g}$, of which we adopt the mean $r_{200} = 1.7$ Mpc -- the mean value of all their sample. Though there are many other observational results in the literature, they are not included in this plot because of (1) the unclear stellar mass cut \citep[e.g.,][]{Guo2012,Shin2021}; (2) inconsistent or unknown halo mass \citep[e.g.,][]{Sales2005,Chen2006, Lares2011, Presotto2012, Hartley2015}.

% Though this value is smaller than the limit ($1.47 \times 10^{10}\ \rm M_{\odot}$) used to constrain stellar mass of galaxies, both limits are within in error range of stellar mass--absolute magnitude relation. 

% Another observation sample from S21 used SZ-selected clusters from the ACT Data Release 5 and galaxies from the Dark Energy Survey Year 3 data set. Their clusters are distributed at $0.15<z<0.7$ with a mean mass of $M_{500} = 3.89 \times 10^{14}\ \rm M_{\odot}$. The absolute magnitude of galaxies, $M_i$, is required to be smaller than $-$19.87. 

In Fig.~\ref{fig:Ngal_prop}, we present the galaxy number density profiles as a function of projected comoving distance with comparisons to the observational results and their redshift evolution.
At $z = 0$, the satellite number density profile shows almost no difference between \GIZ\ and \GX, while vdB15 presents a slightly high profile probably due to our adoption of unified $r_{200}$ for whole cluster sample or the photometric error drawn into SED fitting in vdB15. However, B12 shows a slightly lower profile which could be caused by the redshift evolution or the stellar mass limitation of galaxies. Note that there is a clear redshift evolution as indicated by both simulations. The large gap between the profiles at $z=0$ and $z=1$ also supports the idea that the difference with B12 could be due to redshift evolution. Observed galaxy number density profiles show a cuspy core in vdB15, while simulated profiles show a very flatter curve in the centre of the cluster, as well as B12. 
% This is because of the projection effect??? 
Though there is almost no difference between \GIZ and \GX at $z=0$, the separation between the two runs grows with redshift and \GX produces systematically fewer galaxies than \GIZ at high redshift.
In addition, as redshift increases, the outer radii drop much faster than the inner region, which reflects that galaxies in the centre region of the cluster have assembled at an early time.

\subsection{Stellar metallicity} \label{sec:stellar_metal}

\begin{figure}
    \centering
    \includegraphics[width=0.48\textwidth]{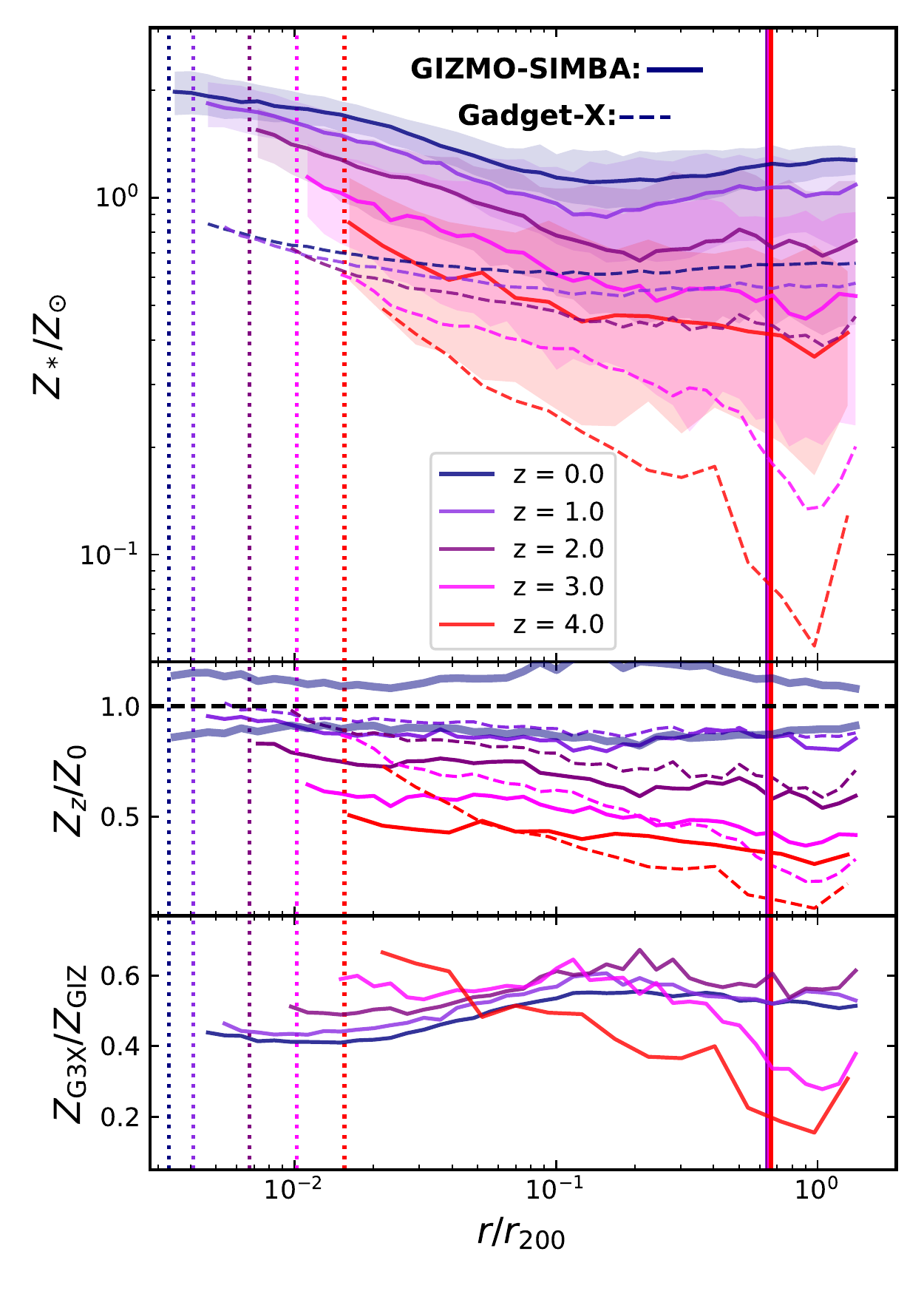}
    \caption{The evolution of the radial profile of mass-weighted stellar metallicity. The radius is re-scaled to $r_{200}$ with the vertical lines showing the median $r_{500}/r_{200}$ in \GIZ. The colours indicate the redshifts as shown in the legend with line styles that have the same meaning as previous plots. The position of softening length at different redshifts is shown with the vertical dotted lines.}
    \label{fig:stemetal}
\end{figure}

The ICM metallicity is rooted in the stellar metallicity from which the metals are disseminated via winds and ram-pressure stripping \citep[e.g.][and references therein]{Holler2014}. Therefore, investigating the stellar metallicity profile will provide the source of the ICM metallicity profile. Furthermore, the stellar metallicity profile may also give clues on the formation history of clusters, inspired by the idea that more metal-rich galaxies/stars should form later. 

In Fig.~\ref{fig:stemetal}, we show the radial profile of mass-weighted stellar metallicity from $z = 4$ to $z = 0$ as a function of radius normalized to $r_{200}$. It is interesting to see that the metallicity profiles from \GIZ\ are systematically higher than that from \GX\ at all redshifts. A roughly two-times difference is shown, though with a slightly different dependence on both redshift and radius. More details can be viewed in the bottom panel of Fig.~\ref{fig:stemetal}. This links to the earlier and higher star formation in \GIZ than \GX \citep[see Fig. 3 in][]{Cui2022}. The earlier and higher star formation in \GIZ\ at high redshift also results in an old BCG age and redder colour compared to \GX. 
The other interesting result is the redshift evolution of metallicity profiles: metallicity is supposed to be cumulative along with time, but there is a less clear redshift evolution in the cluster core region, especially for \GX, which is understandable -- BCG normally forms at first; at large radii, $r \gtrsim 0.2 r_{200}$, both simulations exhibit clear dependence on redshift with lower metallicity at higher redshift. This is expected as there are fewer contributions from satellite galaxies at high redshift than low redshift, which is the main source of metal. This is supported by \cite{Li2020} which showed a similar metallicity radial distribution at $z=0$ by using satellite galaxies. We further note that the weak radial decrement in the metallicity profile at $z=0$, less than a 50 per cent decrease in metallicity compared to the cluster centre region, is in agreement with these observational results of nearby galaxy clusters either with individual metals \citep{Tiwari2020} or galaxy ages \citep{Rakos2007}.

\section{Conclusions} \label{sec:cons}

In \citet{Li2020}, we studied the distributions of the baryonic components of the clusters at $z=0$. As a continuation of studying baryon physical distribution in galaxy clusters, here we focus on the theoretical evolution of gas and stellar physical profiles with two hydrodynamical simulations \GIZ\ and \GX\ from \thethreehundred\ through the tracking of their progenitors. In terms of gas properties, we consider the mass density, electron number density, temperature, entropy, pressure and metallicity profiles, while for the stellar properties, we study the stellar mass, satellite number and metallicity density profiles. We show the radial profiles beyond $r_{500}$ from $z = 4$ to $z = 0$ and include comparisons with observational results. We discuss the effects of different baryon models on the distribution of baryons and meanwhile investigate the self-similarity of profiles. 

Our main conclusions are summarized as follows:

\begin{description}

    \item[{\bf [global evolution]}] Based on different temperature ranges, the gas component is divided into four phases: hot, warm-hot, warm-cold and cold. At high redshift, cold and warm-cold gas is expected to dominate the gas mass fraction but becomes negligible after $z \sim 2-3$. While the fraction of hot gas overtakes all fractions from the other phases at $z \sim 3$. Despite the similarity of these fraction evolution between \GIZ\ and \GX, there are quantitative differences in gas fractions of 4 phases at different redshifts, which indicate the impacts of the two baryon models. 
    % The warm-hot gas only occupies a significant fraction at $z \sim 3-4$.

    \item[{\bf [global evolution]}] The hot and warm-hot gas is almost not bound to galaxies at all redshifts, while almost all warm-cold gas locates in galaxies (also a fraction shown in the diffused component from \GX) and the cold gas can co-exist in both bound and diffuse components at $z \gtrsim 3$. For the stellar components, the mass in galaxies is always higher than that in ICL. Both simulations present a consistent evolution of the abundances of the stellar mass, but more stellar mass in \GX\ is assigned to ICL than that in \GIZ, implying that the stellar distribution outside the galaxy range may be affected by AGN feedback in a different way though we consider a stronger effect of AGN feedback in \GIZ. 

    \item[{\bf [gas profile evolution]}] In the cores of these physical profiles, the two simulations generally exhibit a different shape of profiles for gas properties, while \GX\ is closer to the low-redshift observations. The inner difference should be primarily caused by the AGN feedback models between the two simulations. 

    \item[{\bf [gas profile evolution]}] In the outskirts of physical profiles, a similarity of gas physical properties between two simulations including mass density, electron number density, entropy and pressure appears at low redshift and both are consistent with observations ($r \gtrsim 0.3r_{500}$). Meanwhile, these physical profiles are steeper than that in the cores with a decreasing scatter until $\sim r_{500}$, implying the weak influence of the non-gravitational process at this radius. The self-similarity is generally holding at $z \lesssim 2$, then shows a deviation at high redshifts depending on the simulation.   

    %gas mass density
    \item[{\bf [gas density profile]}] A more cored centre of gas mass density is shown in \GIZ\ than that \GX, with the latter being thus in better agreement with the cuspy profiles indicated by observations. 

    %gas temperature
    \item[{\bf [gas temperature profile]}] For gas temperature, \GX\ shows a low and flat profile in better agreement with the observation data in the core, while \GIZ\ tends to be closer to observation data at intermediate radii. The difference is caused by the kinetic AGN feedback models adopted in \GIZ, i.e., gas particles are ejected by AGN outflow. Meanwhile, a self-similarity holds in \GX\ at all redshift ranges and extends within $0.1r_{500}$, but \GIZ\ presents a redshift evolution. 

    %gas entropy
    \item[{\bf [gas entropy profile]}] In the core region of gas entropy, \GX\ tends to follow a power law slope in line with observational results, while \GIZ\ shows an excess accompanied by a flatter profile. The excess is supposed the result of the high temperature and low-density profiles in \GIZ\, which doesn't contain any cool-core clusters. 

    \item[{\bf [gas pressure profile]}] Albeit much fewer differences in the cluster core region, the two simulations show similar profiles and evolution, which indicate the minimum impacts of the baryon models on the gas pressure profile.
    
    %Gas metallicity
    \item[{\bf [gas metallicity profile]}] A higher gas metallicity profile is shown in \GIZ\ than that in \GX\ at $z=0$, especially at the core where \GX\ agrees with observed ICM metallicity profiles. At $r \gtrsim 0.1r_{500}$, the metallicity decreases with radius in both simulations, while \GIZ\ shows a better agreement with observations but is steeper which should be caused by the differences between mass-weighted and emission-weighted metallicity. There is almost no redshift evolution for \GX\ out to $z \sim 2$, but a clear and gradual redshift dependence is present in \GIZ. We emphasize that it is hard to significantly determine the redshift evolution of metallicity or not, under the large errors of observational results. 

    %Stellar mass density
    \item[{\bf [stellar density profile]}] In the very inner region, \GIZ\ shows a flatter stellar mass density profile compared with \GX\ at $z=0$, which is supposed to be caused by the strong AGN feedback model adopted in \GIZ. At $z \geq 1$, \GIZ\ presents a higher stellar mass density than \GX, which is caused by an early time of consuming more gas into stars in \GIZ, but a strong feedback model quenches galaxy stellar growth finally making a low stellar mass density. The differences in profiles between the two simulations become small with a larger radius. In addition, the slope of the stellar profile becomes steep from low to high redshift. Meanwhile, the stellar mass density profile is highly constrained at $r\sim0.1r_{200}$ despite redshifts and adopted baryon models in simulations, which is also indicated by observation.  

    % Satellite number density
    \item[{\bf [galaxy number density profile]}] The satellite number density profile in two simulations reasonably matches well with low-redshift observations at the outskirts and shows almost no difference between the two simulations at $z=0$, while a separation gradually appears at high redshift. In addition, a much faster drop in number density is shown at the outer radius than that in the inner region, reflecting an early time of assembly history for the galaxies in the centre region of the cluster.

    % stellar metallicity
    \item[{\bf [stellar metallicity profile]}] The stellar metallicity profiles in \GIZ\ are systematically higher than those in \GX\ at all redshifts with an average two times difference. This is related to the earlier and higher star formation in \GIZ\ than \GX. Further, there is a clearer redshift evolution in the outskirts than in the cores due to an early time of BCG formation normally.

    \item[{\bf [fixed-mass evolution]}] Besides the theoretical investigation through tracking progenitors, we also provide fixed-cluster-mass evolution in \autoref{sec:fixedM} which will be useful for comparison with observation results and helps us to understand the mass evolution effects. Reversed evolution trends are found for the fixed-halo-mass profiles of gas density, temperature, and entropy which indicate the mass dependence of these profiles after the normalization. However, the pressure profile agrees with the tracking evolution but with great changes between different redshifts. The gas metallicity profiles are very similar between tracking and fixed-halo mass. In addition, a slight mass dependence of stellar mass density profiles is shown in the centre.
    
\end{description}

To conclude, \GIZ, though calibrated to give a better agreement to the observation in galaxy properties \citep{Cui2022}, tends to have a little higher ICM temperature with relatively low gas density at the cluster centre. This is largely due to the poor resolution in this version of the300 clusters which can not work properly with the SIMBA baryon model. We found the entropy profile of haloes that were simulated with different resolutions showed a clear resolution dependence with the SIMBA model (Cui et al. in prep.). Furthermore, this poor resolution is also responsible for the flat entropy profile at the cluster centre due to numerical reasons. Nevertheless, we find that the high star formation rate in \GIZ\ at high redshift can result in denser stellar cores in galaxies. This can produce a stronger galaxy-galaxy strong lensing signal (Meneghetti et al. in prep.), which points to a possible solution for the tension found in \cite{Meneghetti2020}, of which this tension seemed unable to be solved without producing unrealistic galaxies in the simulation \citep{Meneghetti2022}.
Nevertheless, this paper provides an advanced and comprehensive description of the evolution of baryon physical profiles with redshift and radius changes between the \GIZ\ and \GX, which helps us to understand the pros and cons of these baryon models, thus our knowledge of galaxy cluster evolution. Prospective deep X-ray, SZ, and galaxy observations, possibly using next-generation telescopes, are needed to distinguish between the two models, which exhibit the largest difference in their high-redshift gas and stellar profiles -- a prediction instead of calibration.

\section*{Acknowledgements}
The authors sincerely thank the anonymous referee for the valuable comments.
We gratefully acknowledge the support of the Key Laboratory for Particle Physics, Astrophysics and Cosmology, Ministry of Education. This work is supported by the national science foundation of China (Nos. 11833005, 11890692, 11621303), 111 project No.B20019, and Shanghai Natural Science Foundation, grant No. 19ZR1466800. We acknowledge the science research grants from the China Manned Space Project with Nos. CMS-CSST-2021-A02. WC is supported by the STFC AGP Grant ST/V000594/1 and the Atracci\'{o}n de Talento Contract no. 2020-T1/TIC-19882 granted by the Comunidad de Madrid in Spain. WC and AK thank the Ministerio de Ciencia e Innovación (Spain) for financial support under Project grant PID2021-122603NB-C21. MM acknowledges financial support from PRIN-MIUR 2017WSCC32 and 2020SKSTHZ, from the grant ASI n.2018-23-HH.0. CG, from INAF ``main-stream'' projects 1.05.01.86.20 and 1.05.01.86.31, and from INAF "Supporto alla ricerca di base" (project "The big-data era of cluster lensing"). AK further thanks Sonic Youth for goo. KD acknowledges support by the COMPLEX project from the European Research Council (ERC) under the European Union’s Horizon 2020 research and innovation program grant agreement ERC-2019-AdG 882679 as well as  support by the Deutsche Forschungsgemeinschaft (DFG, German Research Foundation)  under Germany’s Excellence Strategy - EXC-2094 - 390783311. JS was supported by the National Aeronautics and Space Administration under Grant No. 80NSSC18K0920 issued through the ROSES 2017 Astrophysics Data Analysis Program.

This work has been made possible by the \thethreehundred\ collaboration. The project has received financial support from the European Union’s H2020 Marie Skłodowska-Curie Actions grant
number 734374, i.e. the LACEGAL project. 
The simulations used in this paper have been performed in the MareNostrum Supercomputer at the Barcelona Supercomputing Center, thanks to CPU time granted by the Red Espa$\tilde{\rm n}$ola de Supercomputaci$\acute{\rm o}$n. The CosmoSim database used in this paper is a service by the Leibniz-Institute for Astrophysics Potsdam (AIP). The MultiDark database was developed in cooperation with the Spanish MultiDark Consolider Project CSD2009-00064.

This work has made extensive use of the \textsc{python} packages -- \textsc{ipython} with its \textsc{jupyter} notebook \citep{ipython}, \textsc{numpy} \citep{NumPy} and \textsc{scipy} \citep{Scipya,Scipyb}. All the figures in this paper are plotted using the python matplotlib package \citep{Matplotlib}. This research has made use of NASA's Astrophysics Data System and the arXiv preprint server. The computation of this work is partly carried out on the \textsc{Gravity} supercomputer at the Department of Astronomy, Shanghai Jiao Tong University.

%%%%%%%%%%%%%%%%%%%%%%%%%%%%%%%%%%%%%%%%%%%%%%%%%%
\section*{Data Availability}
The data underlying this paper will be shared on reasonable request to the corresponding author.

%%%%%%%%%%%%%%%%%%%% REFERENCES %%%%%%%%%%%%%%%%%%

% The best way to enter references is to use BibTeX:

\bibliographystyle{mnras}
\bibliography{paper} % if your bibtex file is called example.bib

% Alternatively you could enter them by hand, like this:
% This method is tedious and prone to error if you have lots of references
%\begin{thebibliography}{99}
%\bibitem[\protect\citeauthoryear{Author}{2012}]{Author2012}
%Author A.~N., 2013, Journal of Improbable Astronomy, 1, 1
%\bibitem[\protect\citeauthoryear{Others}{2013}]{Others2013}
%Others S., 2012, Journal of Interesting Stuff, 17, 198
%\end{thebibliography}

%%%%%%%%%%%%%%%%%%%%%%%%%%%%%%%%%%%%%%%%%%%%%%%%%%

%%%%%%%%%%%%%%%%% APPENDICES %%%%%%%%%%%%%%%%%%%%%

\appendix

\section{Fixed halo mass evolution} \label{sec:fixedM}
\begin{figure}
    \centering
    \includegraphics[width=0.48\textwidth]{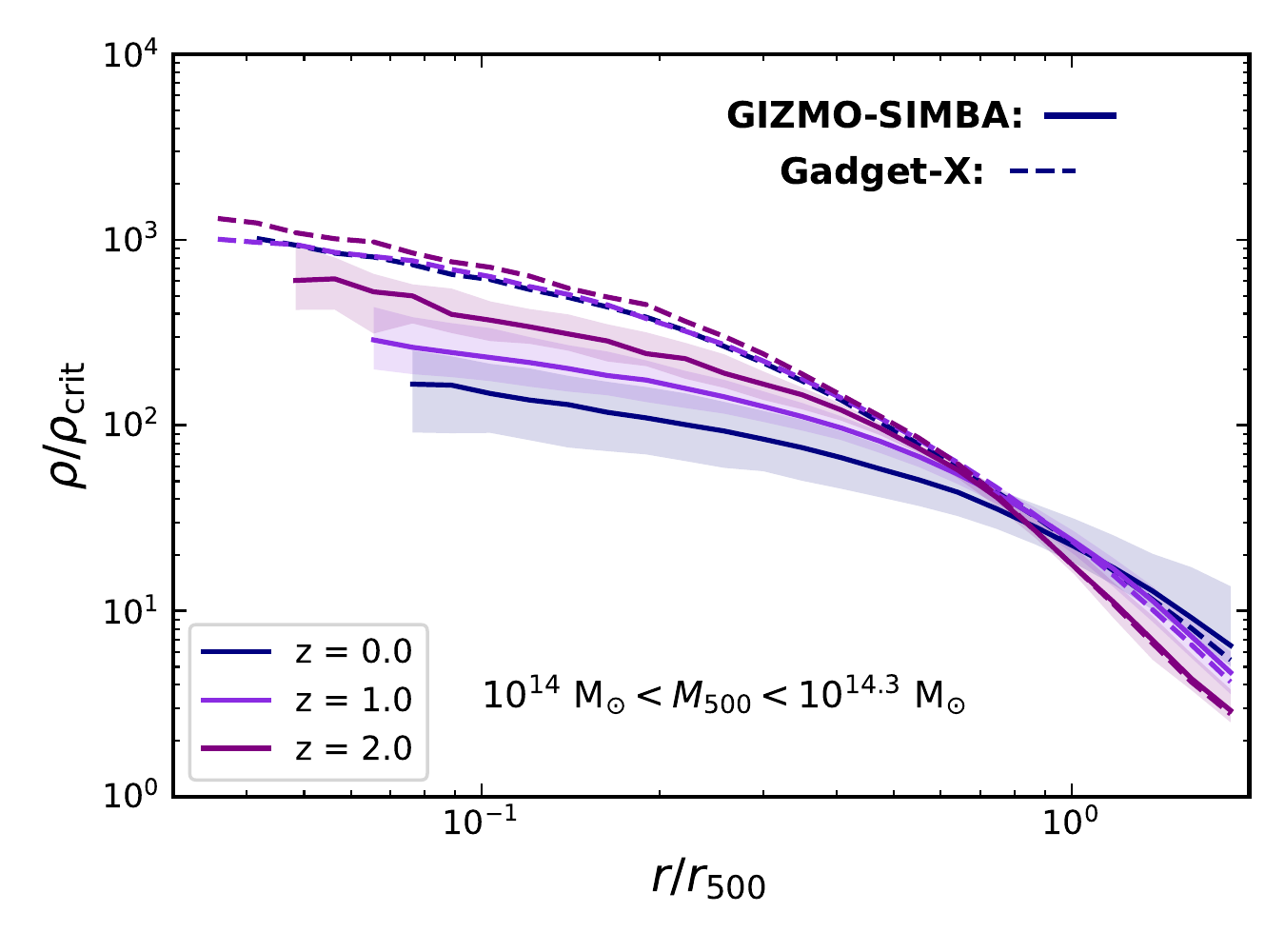}
    \caption{The evolution of gas mass density profiles from $z=2$ to $z=0$ for the haloes with $10^{14}\ \mathrm{M_{\odot}} < M_{500} < 10^{14.3}\ \rm M_{\odot}$. The density is normalized by the critical density of the universe. The solid and dashed lines show the median profiles for \GIZ and \GX, respectively. The shaded regions represent the $16$th and $84$th percentile of all cluster profiles in \GIZ.}
    \label{fig:dens_appd}
\end{figure}

\begin{figure}
    \centering
    \includegraphics[width=0.48\textwidth]{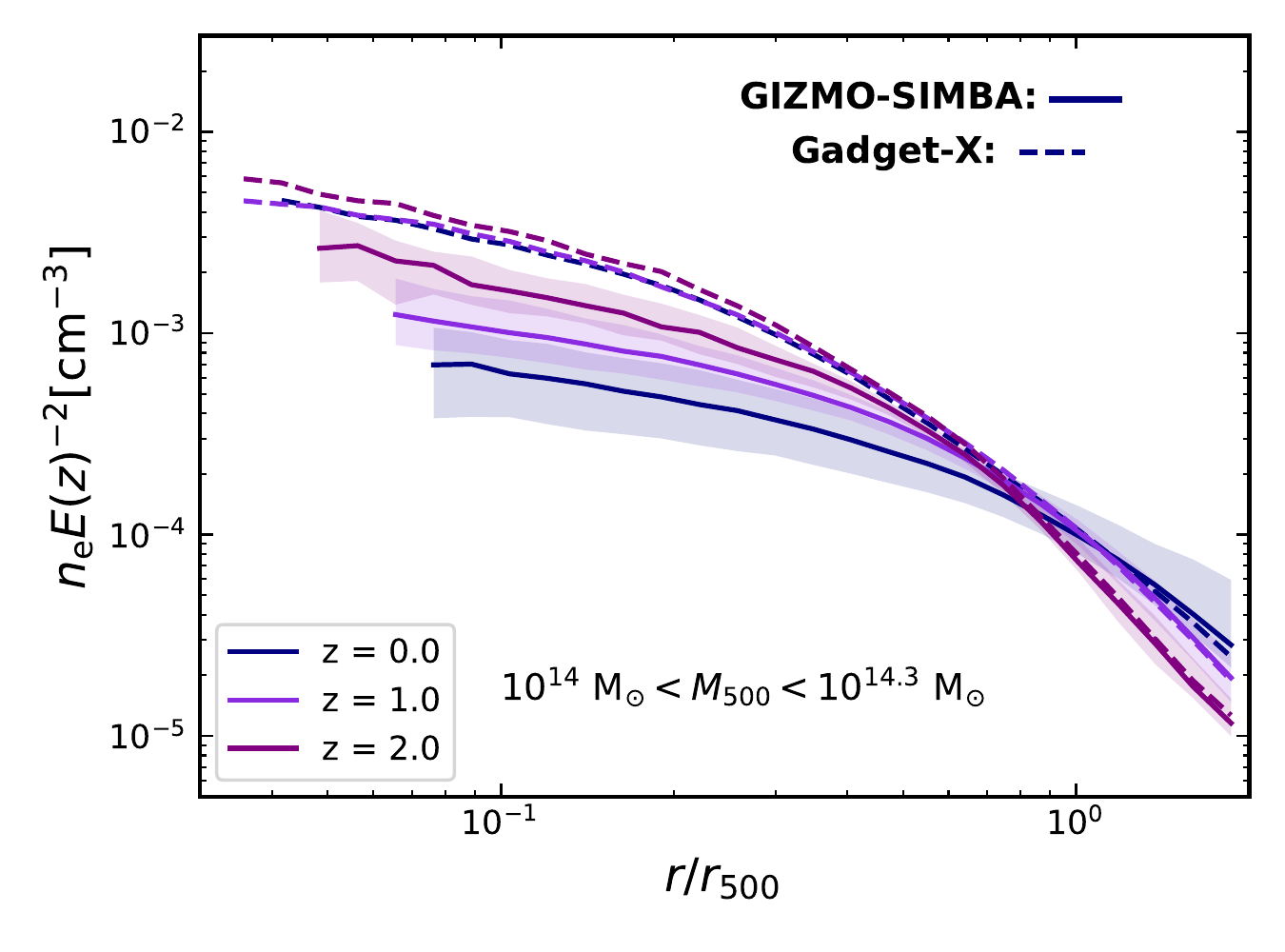}
    \caption{The evolution of electron number density profiles from $z=2$ to $z=0$ for the haloes with $10^{14}\ \mathrm{M_{\odot}} < M_{500} < 10^{14.3}\ \rm M_{\odot}$.}
    \label{fig:elen_appd}
\end{figure}

\begin{figure}
    \centering
    \includegraphics[width=0.48\textwidth]{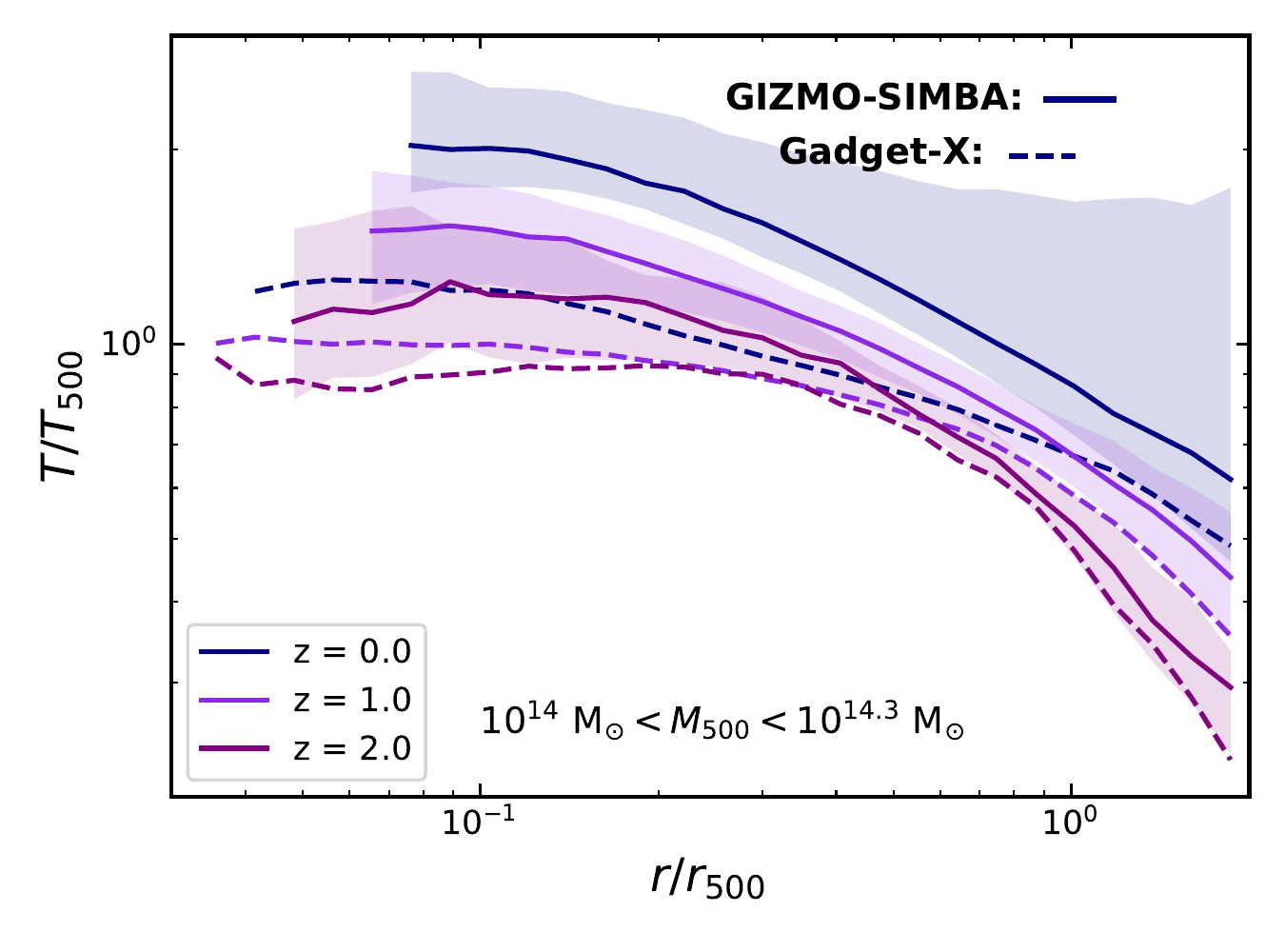}
    \caption{The evolution of mass-weighted temperature profiles from $z=2$ to $z=0$ for the haloes with $10^{14}\ \mathrm{M_{\odot}} < M_{500} < 10^{14.3}\ \rm M_{\odot}$.}
    \label{fig:temp_appd}
\end{figure}

\begin{figure}
    \centering
    \includegraphics[width=0.48\textwidth]{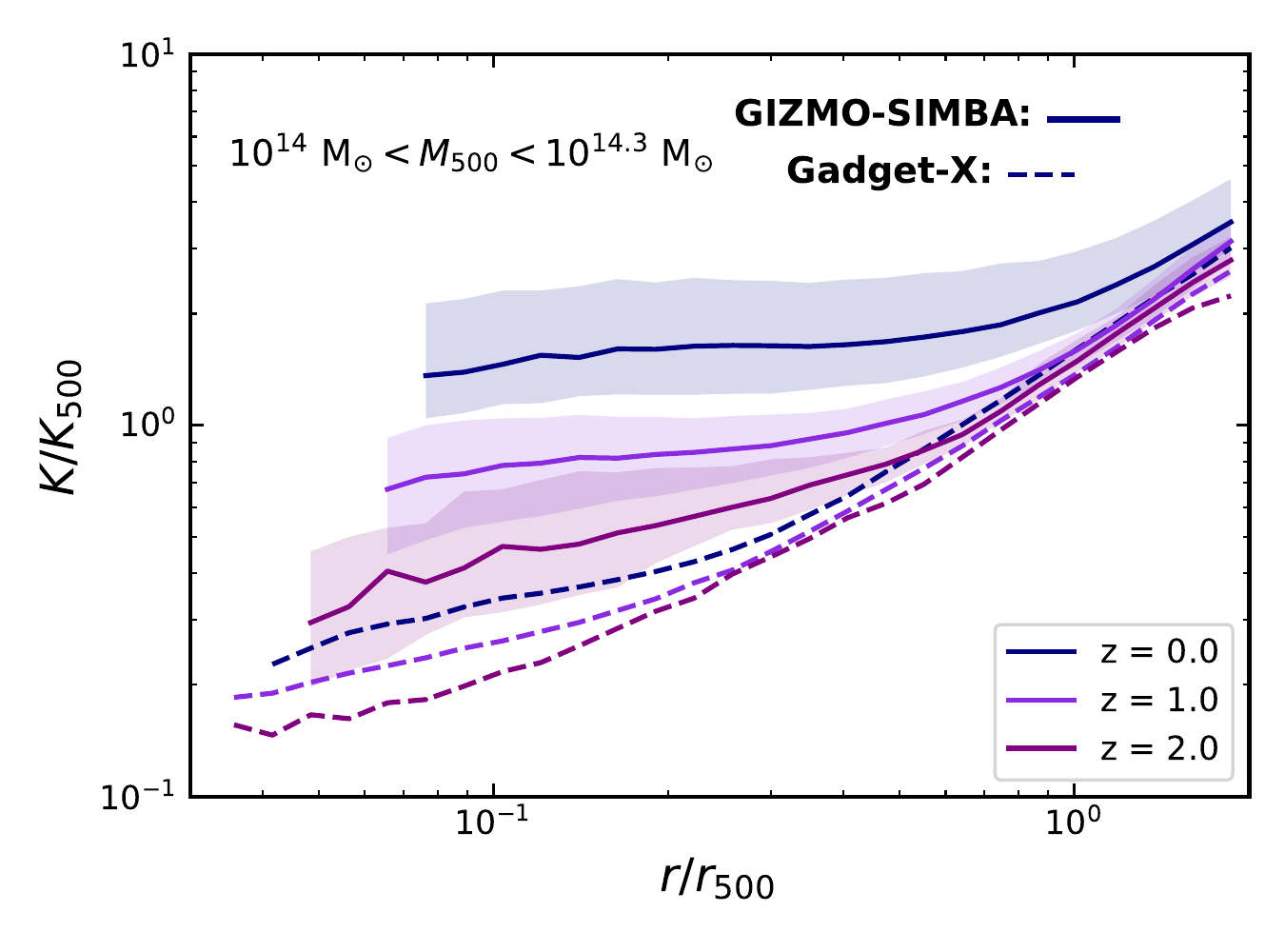}
    \caption{The evolution of entropy profiles from $z=2$ to $z=0$ for the haloes with $10^{14}\ \mathrm{M_{\odot}} < M_{500} < 10^{14.3}\ \rm M_{\odot}$.}
    \label{fig:entropy_appd}
\end{figure}

\begin{figure}
    \centering
    \includegraphics[width=0.48\textwidth]{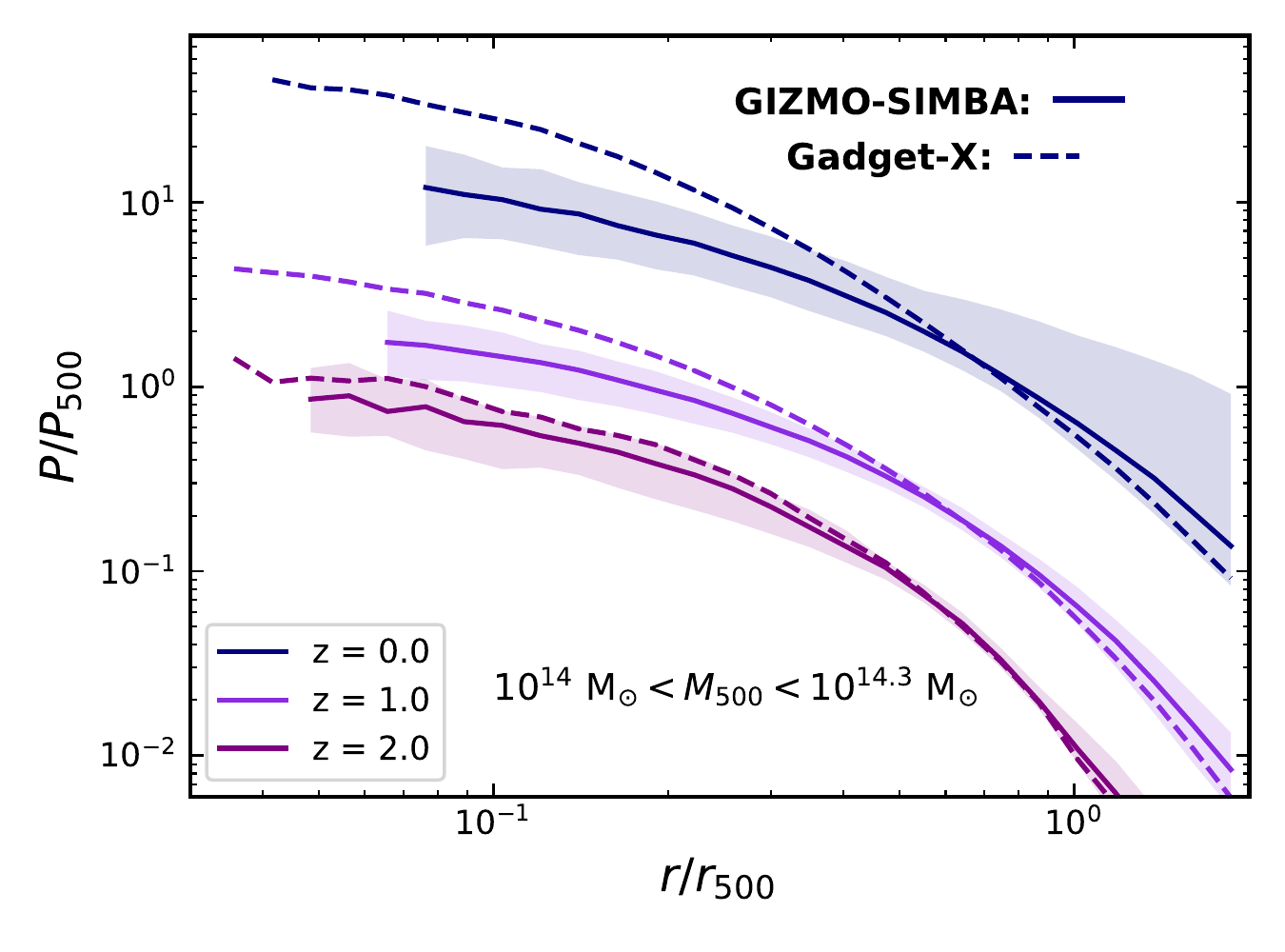}
    \caption{The evolution of pressure profiles from $z=2$ to $z=0$ for the haloes with $10^{14} \mathrm{M_{\odot}}\ < M_{500} < 10^{14.3}\ \rm M_{\odot}$.}
    \label{fig:pres_appd}
\end{figure}

\begin{figure}
    \centering
    \includegraphics[width=0.48\textwidth]{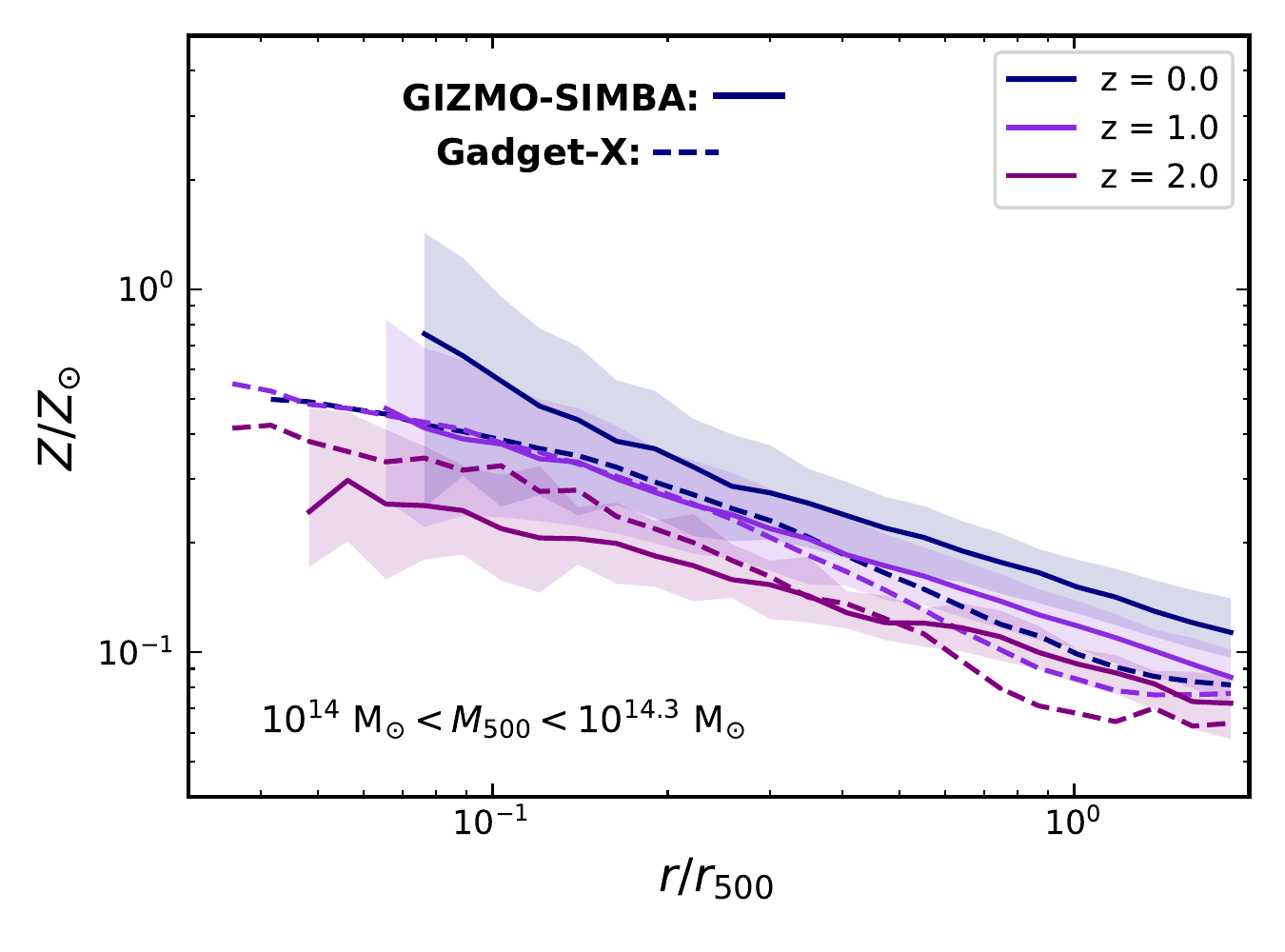}
    \caption{The evolution of metallicity profiles from $z=2$ to $z=0$ for the haloes with $10^{14}\ \mathrm{M_{\odot}} < M_{500} < 10^{14.3}\ \rm M_{\odot}$.}
    \label{fig:metal_appd}
\end{figure}

We investigate the mass dependence of the evolution of gas profiles by adopting a fixed halo mass range of $10^{14}\ \mathrm{M_{\odot}} < M_{500} < 10^{14.3}\ \rm M_{\odot}$ in this section. We show the evolution of gas mass density, electron number density, mass-weighted temperature, entropy, pressure and metallicity from $z = 2$ to $z = 0$ in Fig.~\ref{fig:dens_appd}, Fig.~\ref{fig:elen_appd}, Fig.~\ref{fig:temp_appd}, Fig.~\ref{fig:entropy_appd}, Fig.~\ref{fig:pres_appd}, and Fig.~\ref{fig:metal_appd}, respectively. We run out of enough samples at $z=3$ due to the small region of the zoomed simulation, even with this low mass cluster mass bin. Thus, the comparison can only be done up to $z=2$.

In inner radii, the redshift evolution of gas mass density profiles behaves diversely between \GX\ and \GIZ. There is almost no redshift evolution in \GX, which reflects that the deviation from self-similarity shown in Fig.~\ref{fig:Gdens} up to $z=2$ is caused by the mass evolution. While \GIZ\ shows a clear reversed evolution compared with the density evolution of the progenitors of clusters (Fig.~\ref{fig:Gdens}), indicating a stronger mass dependence than \GX. The gas mass density profiles at outer radii decrease with redshift increasing instead and are consistent between \GX\ and \GIZ, which illustrates that the amount of gas is less sensitive to both the baryon models and halo masses. The electron number density profiles show the same performance as gas mass density profiles. The temperature profiles also present a contrary tendency to the profiles shown in Fig.~\ref{fig:MWTemp}. At $z = 0$, the haloes with $10^{14}\ \mathrm{M_{\odot}} < M_{500} < 10^{14.3}\ \rm M_{\odot}$ have a higher temperature than clusters. Similar to the gas density profile, this redshift evolution indicates halo mass is not the only key to determining the temperature profiles, again much stronger in \GIZ\ than \GX.
% These less massive haloes are expected to form at an early time and are more relaxed compared with massive clusters, thus having a high temperature \citep[see the comparison of temperature between relaxed and unrelaxed clusters in][]{Li2020}. 
The reversed evolution of entropy profiles contrasted to Fig.~\ref{fig:Entropy} also reflects a mass-dependent deviation from self-similarity in the inner region of clusters. While the mass dependence becomes weak at the outskirts. The pressure profiles follow the evolution of tendency in Fig.~\ref{fig:Pressure} despite a large separation, which implies an increment along with cluster thermalization. The gas metallicity no doubt increases with a reduction of redshift as a result of the cumulative metallicity effect. Compared with the amplitude of the profile for massive clusters, the evolution of metallicity is slightly influenced by halo mass.

\begin{figure}
    \centering
    \includegraphics[width=0.48\textwidth]{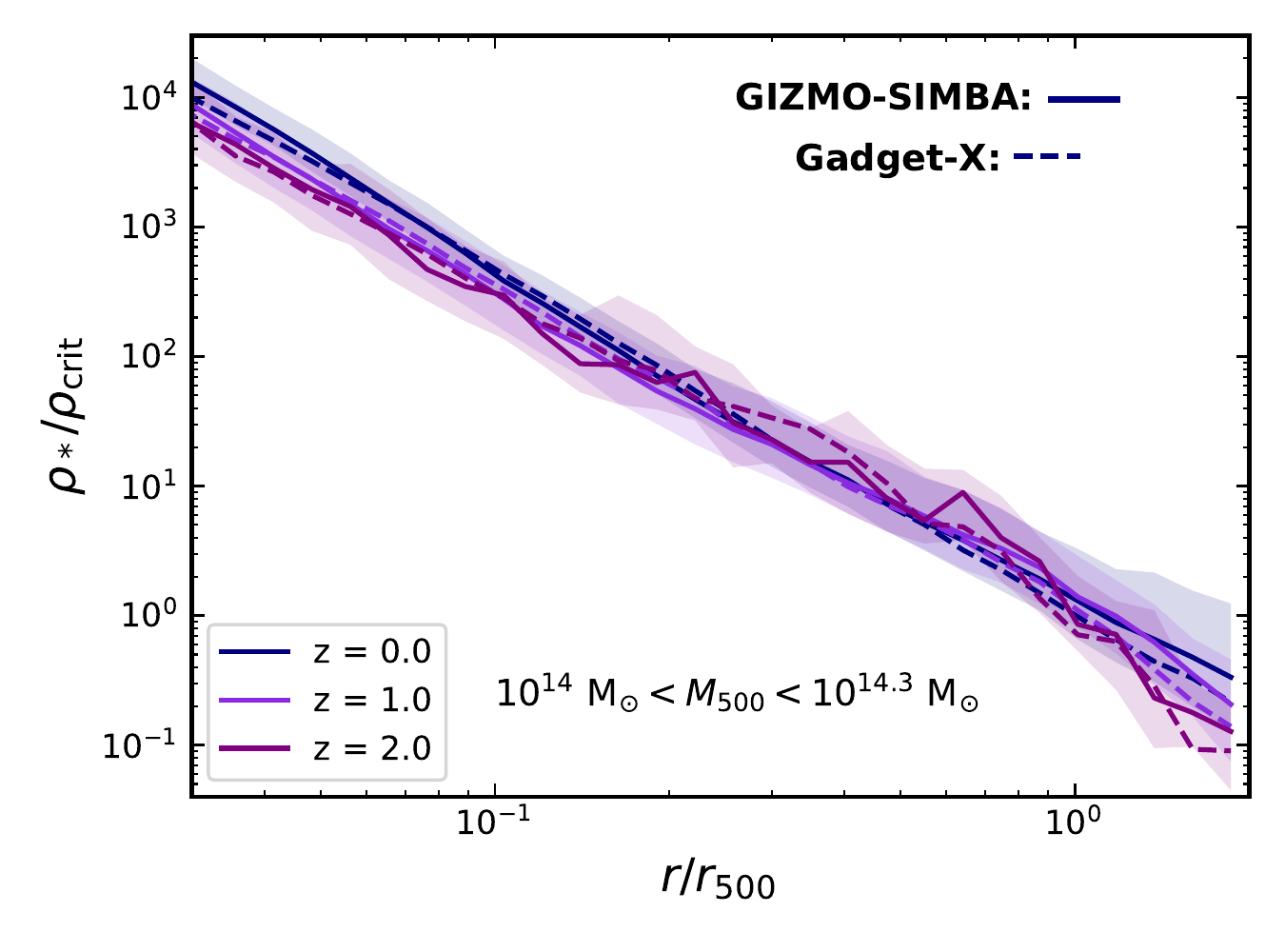}
    \caption{The evolution of stellar mass density profiles from $z=2$ to $z=0$ for the haloes with $10^{14}\ \mathrm{M_{\odot}} < M_{500} < 10^{14.3}\ \rm M_{\odot}$.}
    \label{fig:stedens_appd}
\end{figure}

In addition, we show the stellar mass density profiles at fixed halo mass in Fig~\ref{fig:stedens_appd}. The stellar mass density shows a power-law distribution independent of redshift and baryon models, accompanied by a slightly decreasing tendency with redshift increasing.

\section{Total mass density}

\begin{figure}
    \centering
    \includegraphics[width=0.48\textwidth]{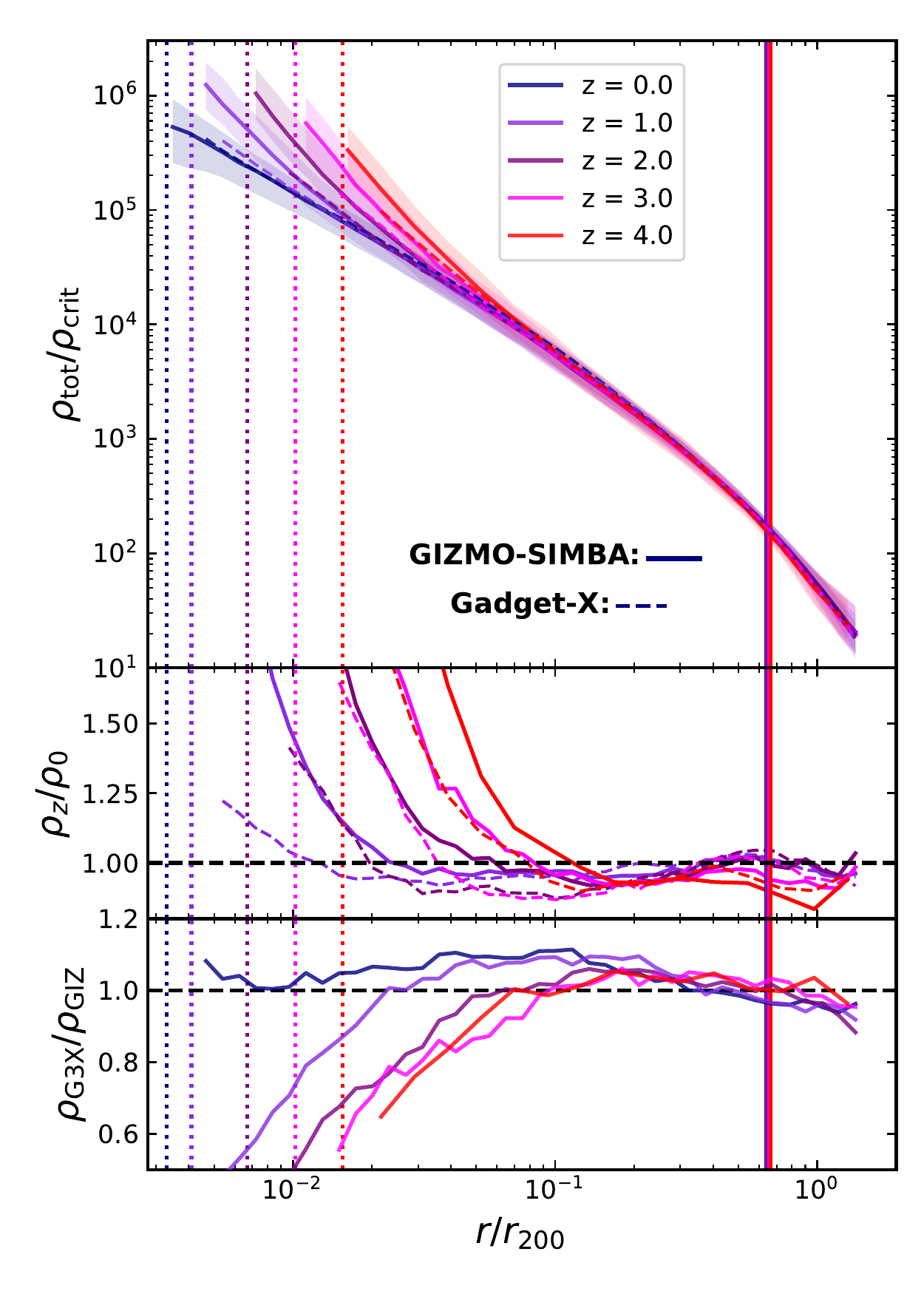}
    \caption{The evolution of total mass density scaled with the critical density of the universe.}
    \label{fig:Totdens}
\end{figure}

We present the total mass density profiles in Fig.~\ref{fig:Totdens}. The total mass density profiles show a high self-similarity across all the redshift ranges at $r>0.1r_{500}$ in both \GIZ\ and \GX, where the amplitude is also highly consistent with each other. This is because the major component of galaxy clusters, DM, not involved with baryonic effects, dominates the matter distribution at this radius. 

\bsp	% typesetting comment
\label{lastpage}
\end{document}